\documentclass[11pt]{article}

\usepackage[OT2,OT1]{fontenc}

\usepackage{a4wide}
\usepackage{amsmath}

\usepackage{mathrsfs}
\usepackage[T1]{fontenc}
\usepackage{mathpazo}
\usepackage{setspace}
\usepackage{amsfonts}
\usepackage{amssymb}
\usepackage{amsmath}
\usepackage{epsfig}
\usepackage{latexsym}
\usepackage{color}
\usepackage{nicefrac}
\usepackage[latin1]{inputenc}
\usepackage{cite}

\numberwithin{equation}{section}

\definecolor{papercolor}{rgb}{0.25, 0 ,1}
\definecolor{markcolor}{rgb}{1, 0.2, 0.2}

\newcommand{\oao}[2]{{#1\atopwithdelims[]#2}}

\definecolor{fux}{rgb}{0.75, 0 ,1}

\author{
  \begin{minipage}{.97\linewidth}
    \vspace{1cm}
       \begin{center}
      \begin{small}
        \textbf{P. Marios Petropoulos}${\, }^1$ and
                       \textbf{Pierre Vanhove}${\,}^2$        
              \end{small}
    \end{center}
    \vspace{0.5cm}
    \hspace{2cm}\begin{minipage}{.7\linewidth}
\begin{center}     {\it \begin{footnotesize}
\hbox{\vbox{
 \begin{itemize}
              \item[${}^1$] Centre de Physique Th\'eorique\\ 
        Ecole Polytechnique, CNRS UMR 7644\\
        91128 Palaiseau Cedex, France\\
{\tt marios@cpht.polytechnique.fr}
\end{itemize}\vskip1cm}
\kern-5cm\vbox{
\begin{itemize}
         \item[${}^2$] Institut des Hautes Etudes Scientifiques\\ 
        Le Bois-Marie\\
        91440 Bures-sur-Yvette, France\\
        and\\    
        Institut de Physique Th\'eorique\\ 
        CEA, CNRS URA 2306\\
        91191 Gif-sur-Yvette, France \\
{\tt pierre.vanhove@cea.fr}   \end{itemize}
}}
     \end{footnotesize}}
\end{center}
    \end{minipage}
    \vspace{0.5cm}
  \end{minipage}
}

\title{\vspace{1.5cm}
 \boldmath \begin{huge}
    \textbf{Gravity, strings, modular 
and quasimodular~forms}
  \end{huge} \unboldmath
}

\begin{document}

\begin{titlepage}
  \maketitle
  \thispagestyle{empty}

  \vspace{-12.5cm}
  \begin{flushright}
    CPHT-RR005.0211\\
IPHT-t12/016\\
IHES/P/12/08
  \end{flushright}

  \vspace{10cm}

  \begin{center}
    \textsc{Abstract}\\
  \end{center}
Modular and quasimodular forms have played an important role in gravity and string theory. Eisenstein series  have appeared systematically in the determination of spectrums and partition functions, in the description of non-perturbative effects, in higher-order corrections of scalar-field spaces, \dots The latter often appear as gravitational instantons \emph{i.e.} as special solutions of Einstein's equations. In the present lecture notes we present a class of such solutions in four dimensions, obtained by requiring (conformal) self-duality and Bianchi IX homogeneity. In this case, a vast range of configurations exist, which exhibit interesting modular properties.  Examples of other Einstein spaces, without Bianchi IX symmetry, but with similar features are also given. Finally we discuss the emergence and the role of Eisenstein series in the framework of field and string theory perturbative expansions, and motivate the need for unravelling novel modular structures.

\vspace{3cm} \noindent To appear in the proceedings of the
\textsl{Besse Summer School on Quasimodular Forms -- 2010}.

\end{titlepage}

\onehalfspace

\tableofcontents

\section{Introduction}

Modular forms often appear in physics as a consequence of duality properties. This comes either as an invariance of a theory or as a relationship among two different theories, under some discrete transformation of the parameters. The latter transformation can be a simple  $\mathbb{Z}_2$ involution or an element of some larger group like $SL(2,\mathbb{Z})$. The examples are numerous and have led to important developments in statistical mechanics, field theory, gravity or strings. 

One of the very first examples, encountered in the 19th century, is the electrtic--magnetic duality in vacuum Maxwell's equations (see e.g. \cite{J99}), which are invariant under interchanging electric and magnetic fields. This was revived in the more general framework of Abelian gauge theories by Montonen and Olive in 1977 \cite{MO77} and culminated in the Seiberg--Witten duality in supersymmetric non-Abelian gauge theories \cite{SW94}. There  the duality group $SL(2,\mathbb{Z})$ acts on a complex
parameter $ \tau=\frac{\theta}{2\pi} +\frac{4\pi i }{g^2}$, where $\theta$ is the vacuum angle and $g$ is the coupling constant. The modular transformations give thus access to the non-perturbative regime of the field theory. 

In statistical mechanics, the Kramers--Wannier duality \cite{KW41} predicted in 1941 the existence of a critical temperature $T_c$ separating the ferromagnetic ($T<T_c$) and the paramagnetic  ($T>T_c$) phases in the two-dimensional Ising model. The canonical partition function of the model, computed  a few years later by Lars Onsager \cite{ON44}, is indeed expressed in terms of modular forms. 

Over  the last 30 years, modular and quasimodular forms have mostly
emerged in the framework of gravity and string theory. At the first
place, one finds (see e.g. \cite{GSW87}) the canonical partition
function of a string of fundamental frequency $\omega$ at temperature
$T$:  
\begin{equation}
  Z=\frac{1}{\eta(q)},    
\end{equation}
where $q=\exp 2i\pi z = \exp -\nicefrac{\hbar \omega}{kT} $ and
$\eta(q)$ the Dedekind function (we refer to the appendix 
for definitions and conventions on theta functions, \eqref{A1}--\eqref{e:useful}). The average energy stored in a string at temperature $T$ is thus
\begin{equation}
\langle E \rangle=
-\frac{\partial \ln Z}{\partial \nicefrac{1}{kT}}=\frac{\hbar \omega}{24} E_2(z),
\end{equation}
where $E_2(z)$  the weight-two quasimodular form. Again, the modular properties of these functions translate into a low-temperature/high-temperature duality, which exhibits a critical temperature, signature of the Hagedorn transition.
  
In the above examples the modular group acts on a modular parameter related to the temperature. There is a plethora of examples in gravity and string theory of more geometrical nature, related to gravitational configurations and in particular to instantons.
  
An instanton is a solution of non-linear field equations resulting  from an imaginary-time \emph{i.e.}  Euclidean action $S[\phi]$ as
\begin{equation}
 \frac{\delta S}{\delta \phi} = 0.
 \end{equation}
It should not to be confused with a soliton. The latter  is a finite-energy solution of non-linear real-time equations of motion and appear in a large palette of phenomena such as the propagation of solitary waves in liquid media (as e.g. tsunamis\footnote{A valuable account of these properties can be found in \cite{OB80}.}) or black holes in gravitational set ups. 

Instantons have finite action and enter the description of quantum-mechanical processes, which are not captured by perturbative expansions, as their magnitude is controlled by 
$\exp -\nicefrac{1}{g}$ at small coupling $g$ (in electrodynamics $g=\nicefrac{e^2}{\hbar c}$). These phenomena include quantum-mechanical tunneling and, more generally,  decay and creation of bound states. Their amplitude is weighted by 
$
\exp( -\nicefrac{S}{\hbar})
$, where $S$ is the action of the instanton solution interpolating between initial and final configurations (see \cite{CE85} for a pedagogical presentation of these methods).

All interacting (\emph{i.e.} non-linear) field theories exhibit instantons. These emerged originally  in Yang--Mills theories \cite{BPST75, TH76} as well as in 
general relativity \cite{Taub-nut, Eguchi:1978gw, Eguchi:1979yx}. In the latter case, their usefulness for the description of quantum transitions is tempered by quantum inconsistencies of general relativity. Such configurations turn out nevertheless to be instrumental in modern theories of gravity, supergravity and strings for at least two reasons. 

At the first place, some gravitational instantons falling in the class
of asymptotically locally Euclidean (ALE) spaces have the required
properties to  serve as compactification set ups for superstring
models usually defined in space--time of  dimensions 10. This is the case, for example, of the Eguchi--Hanson gravitational instanton \cite{Eguchi:1978gw, Eguchi:1979yx}, 
which appears as a blow-up of the $\mathbb{C}_2/\mathbb{Z}_2$ $A_1$-type singularity, or of more general Gibbons--Hawking multi-instantons \cite{Gibbons:1979zt}.

The second reason is that supergravity and string theories contain many scalar fields called moduli. Their dynamics is often encapsulated in non-linear sigma models, which happen to have as a target space certain gravitational instantons such as the Taub-NUT, Atiyah--Hitchin, Fubini--Study, Pedersen or Calderbank--Pedersen spaces \cite{Taub-nut, Atiyah1, GP78, P85, P86, PS88, CP00, CP02}. Due to some remarkable underlying duality properties, most of the spaces at hand are expressed in terms of (quasi) modular forms, and this makes them relevant in the present context. 

It should be finally stressed that in the framework of string and supergravity theories, quasimodular forms do not appear exclusively via compactification or moduli spaces. Recent developments on the perturbative expansions in quantum field theory reveal how relevant the spaces of quasimodular forms are for understanding the ultraviolet behaviour and its connections with string theory acting as a ultraviolet regulator\cite{Green:2008uj}. They also call for introducing new objects, which stand beyond the realm of Eisenstein series \cite{Green:2010kv}. 

\section{Solving Einstein's equations}

It is a hard task in general to solve Einstein's equations. In four dimensions with Euclidean signature, following the paradigm of Yang--Mills, the requirement of self-duality (or of a conformal variation of it) often leads to integrable equations. Those are in most cases related to self-dual Yang--Mills reductions, and possess remarkable solutions (see. e.g. \cite{Ward85}). It should be mentioned for completeness that self-duality can also serve as a tool in more than four dimensions. In seven or eight dimensions, it can be implemented using $G_2$ or quaternionic algebras \cite{Acharya:1996tw, Floratos:1998ba, Bakas:1998rt,Brecher:1999xf}.
It is not clear, at present, whether in those cases some interesting and non-trivial relationship with quasimodular forms emerge. We will therefore not pursue this direction here.

\subsection{Curvature decomposition in four dimensions}

The Cahen--Debever--Defrise decomposition, more commonly known as Atiyah--Hitchin--Singer
\cite{CDD, AHS}, is a convenient taming of the 20 independent components of the Riemann tensor. In Cartan's formalism, these are  captured by a set of curvature two-forms ($a, b, \ldots=1, \ldots,4$)
\begin{equation}\label{curv}
 \mathcal{R}^a_{\hphantom{a}b}=\mathrm{d} \omega^a_{\hphantom{a}b}
 +\omega^a_{\hphantom{a}c}\wedge \omega^c_{\hphantom{c}b} = \frac{1}{2}
 R^a_{\hphantom{a}bcd} \theta^c\wedge \theta^d,
\end{equation}
where $\left\{\theta^a\right\}$ are a basis of the cotangent space and $   \omega^a_{\hphantom{a}b}=\Gamma^a_{\hphantom{a}bc}\theta^c$ the set of connection one-forms obeying the requirement of vanishing torsion
\begin{equation}
\mathcal{ T}^a=\mathrm{d}\theta^a + \omega^a_{\hphantom{a}b}\wedge \theta^b= \frac{1}{2}
    T^a_{\hphantom{a}bc} \theta^b\wedge \theta^c=0.\label{torsles}
\end{equation}
The cyclic and Bianchi identities ($\mathrm{d}\wedge \mathrm{d}  \theta^a=\mathrm{d}\wedge \mathrm{d}  \omega^a_{\hphantom{a}b} =0$), assuming a torsionless connection,  read:
\begin{eqnarray}
\mathcal{R}^a_{\hphantom{a}b} \wedge
    \theta^b&=&0,\\
    \mathrm{d}\mathcal{R}^a_{\hphantom{a}b}
    + \omega^a_{\hphantom{a}c} \wedge \mathcal{R}^c_{\hphantom{c}b} - \mathcal{R}^a_{\hphantom{a}c} \wedge
    \omega^c_{\hphantom{c}b} &=&0.
\end{eqnarray}
We will assume the basis $\{\theta^a\}$ to be orthonormal with respect to the metric $g$
\begin{equation}
 g = \delta_{ab} \theta^a \theta^b,
\end{equation}
and the connection to be metric ($\nabla g =0$), which is equivalent to
\begin{equation}\label{metcon}
 \omega_{ab}=-\omega_{ba}.
\end{equation}
The latter together with (\ref{torsles}) determine the connection.

The general holonomy group in four dimensions  is $SO(4)$, and $g$ is invariant under local transfromations $\Lambda(x)$ such that
\begin{equation}
\theta^{a\prime}= 
\Lambda^{-1\, a}_{\hphantom{-1\, a}b}\theta^{b},
\end{equation}
under which the connection and curvature forms transform as  
 \begin{eqnarray}
 \omega^{a\prime}_{\hphantom{a}b}&=&\Lambda^{-1\, a}_{\hphantom{-1\, a}c}
\omega^{c}_{\hphantom{c}d}
\Lambda^d_{\hphantom{d}b}+ \Lambda^{-1\, a}_{\hphantom{-1\, a}c} \mathrm{d}\Lambda^c_{\hphantom{c}b},\\
\mathcal{R}^{a\prime}_{\hphantom{a}b}&=&
\Lambda^{-1\, a}_{\hphantom{-1\, a}c}
\mathcal{R}^{c}_{\hphantom{c}d}
\Lambda^d_{\hphantom{d}b}.
\end{eqnarray}
Both $\omega_{ab}$ and $\mathcal{R}_{ab}$ are antisymmetric-matrix-valued one-forms, belonging to the representation $\mathbf{6}$ of $SO(4)$.

Four dimensions is a  special case as $SO(4)$ is factorized into $SO(3)\times SO(3)$. Both connection and curvature forms are therefore reduced with respect to each $SO(3)$ factor as
$\mathbf{3}\times \mathbf{1} +\mathbf{1}\times \mathbf{3}$, where $\mathbf{3}$ and $\mathbf{1}$ are respectively the vector and singlet representations ($i,j,\ldots = 1,2,3$):
\begin{eqnarray}
\Sigma_i=\frac{1}{2}\left(\omega_{0i} + \frac{1}{2} \epsilon_{ijk}\omega^{jk}\right),&\quad&
A_i=\frac{1}{2}\left(\omega_{0i} - \frac{1}{2} \epsilon_{ijk}\omega^{jk}\right),\label{sdcon}\\ 
\mathcal{S}_i=\frac{1}{2}\left(\mathcal{R}_{0i} + \frac{1}{2} \epsilon_{ijk}\mathcal{R}^{jk}\right),&\quad&
\mathcal{A}_i=\frac{1}{2}\left(\mathcal{R}_{0i} - \frac{1}{2} \epsilon_{ijk}\mathcal{R}^{jk}\right),
\end{eqnarray}
while (\ref{curv}) reads:
\begin{equation}
 \mathcal{S}_i= \mathrm{d} \Sigma_i -\epsilon_{ijk} \Sigma^j \wedge  \Sigma^k, \quad
  \mathcal{A}_i= \mathrm{d} A_i +\epsilon_{ijk} A^j \wedge A^k.
    \end{equation}
    
   Usually ($\Sigma_i, \mathcal{S}_i$) and ($A_i,  \mathcal{A}_i$) are referred to as self-dual and anti-self-dual components of the connection and Riemann curvature. This follows from the definition of the dual forms (supported by the fully antisymmetric symbol $\epsilon_{abcd}$\footnote{A remark is in order here for $D=7$ and $8$.
The octonionic structure constants  $\psi_{\alpha\beta\gamma}\ \alpha,\beta,\gamma\in\{1,\ldots,7\}$ and the dual $G_2$-invariant antisymmetric symbol $\psi^{\alpha\beta\gamma\delta}$ allow to define a duality relation in 7 and 8 dimensions with respect to an $SO(7)\supset G_2$, and an
$SO(8)\supset \mathrm{Spin}_7$ respectively. Note, however, that neither $SO(7)$ nor $SO(8)$ is factorized, as opposed to $SO(4)$.})
\begin{eqnarray}
\tilde{\omega}^a_{\hphantom{a}b} &=& \frac{1}{2}
    \epsilon^{a\hphantom{bc}d}_{\hphantom{a}bc}\omega^c_{\hphantom{c}d},\\ 
    \tilde{\mathcal{R}}^a_{\hphantom{a}b} &=& \frac{1}{2}
    \epsilon^{a\hphantom{bc}d}_{\hphantom{a}bc}\mathcal{R}^c_{\hphantom{c}d},
\end{eqnarray}
borrowed from the Yang--Mills\footnote{Note the action of the duality on the components, as 
$ \tilde{\omega}^a_{\hphantom{a}b} =\tilde{\Gamma}^a_{\hphantom{a}bc}\theta^c,    \tilde{\mathcal{R}}^a_{\hphantom{a}b}=\frac{1}{2}
    \tilde{R}^a_{\hphantom{a}bcd} \theta^c\wedge \theta^d$:
$$
   \begin{array}{rcl}
    \tilde{\Gamma}^a_{\hphantom{a}bc}&=& \frac{1}{2}\epsilon^{a\hphantom{be}f}_{\hphantom{a}be}\Gamma^{e}_{\hphantom{e}fc},\\
    \tilde{R}^a_{\hphantom{a}bcd}&=& \frac{1}{2}\epsilon^{a\hphantom{be}f}_{\hphantom{a}be}R^{e}_{\hphantom{e}fcd},
    \end{array}
$$
and similarly for the Weyl part or the Riemann.}. Under this involutive operation, ($\Sigma_i, \mathcal{S}_i$) remain unaltered whereas ($A_i,  \mathcal{A}_i$) change sign.
        
Following the previous reduction pattern, the basis of 6 independent two-forms  can be decomposed in terms of two sets of singlets/vectors with respect to the two $SO(3)$ factors:
\begin{eqnarray}
\phi^i&=&\theta^0\wedge\theta^i + \frac{1}{2} \epsilon^i_{\hphantom{i}jk}\theta^j\wedge\theta^k, \label{2fs}\\ 
\chi^i&=&\theta^0\wedge\theta^i - \frac{1}{2} \epsilon^i_{\hphantom{i}jk}\theta^j\wedge\theta^k.
\label{2fa}\end{eqnarray}
In this basis, the 6 curvature two-forms $\mathcal{S}_i$ and $\mathcal{A}_i$ are decomposed as
\begin{equation}
      \begin{pmatrix}
   \mathcal{S}\\ 
   \mathcal{A}
  \end{pmatrix}=    
  \frac{r}{2}  
   \begin{pmatrix}
    \phi\\
   \chi
  \end{pmatrix} ,   
    \end{equation}
where the $6\times 6$ matrix $r$ reads:    
\begin{equation}
r= 
\begin{pmatrix}
    A&C^+\\
   C^-&B
  \end{pmatrix}    =
   \begin{pmatrix}
    W^+&C^+\\
   C^-&W^-
  \end{pmatrix} +\frac{s}{6}\, \mathbf{I}_6.
    \end{equation}

The 20 independent components of the Riemann tensor are stored inside the symmetric matrix  $r$ as follows:
\begin{itemize}
\item $s=\mathrm{Tr}\, r =2 \mathrm{Tr}\, A=2\mathrm{Tr}\, B=\nicefrac{R}{2}$ is the scalar curvature.  
\item  The  9 components of the traceless part of the Ricci tensor $S_{ab}=R_{ab}-\frac{R}{4}g_{ab}$ ($R_{ab}=R^{c}_{\hphantom{a}acb}$) are given in $C^+=\left(C^-\right)^{\mathrm{t}}$ as
\begin{equation}
S_{00}=\mathrm{Tr}\, C^+, 
\quad 
S_{0i}= 
\epsilon_i^{\hphantom{i}jk}C^-_{jk},
\quad 
S_{ij}= C^+_{ij}+C^-_{ij}-\mathrm{Tr}\, C^+ \delta_{ij}.
    \end{equation}
\item The 5 entries of the symmetric and traceless  $W^+$  are the components of the self-dual Weyl tensor, while $W^-$ provides the corresponding 5 anti-self-dual ones.
\end{itemize}
 In summary,    
\begin{eqnarray}
 \mathcal{S}_i&=&\mathcal{W}_i^++\frac{1}{12}s\phi_i+\frac{1}{2}C^+_{ij}\chi^j,\label{scurv}\\ 
   \mathcal{A}_i&=&\mathcal{W}_i^-+\frac{1}{12}s\chi_i+\frac{1}{2}C^-_{ij}\phi^j,\label{acurv} 
\end{eqnarray}
where 
\begin{equation}
\mathcal{W}_i^+=\frac{1}{2}W^+_{ij}\phi^j, \quad
\mathcal{W}_i^-
=\frac{1}{2}W^-_{ij}\chi^j
 \end{equation}
are the self-dual and anti-self-dual Weyl two-forms respectively.
 
Given the above decomposition, some remarkable geometries  emerge (see e.g. \cite{Eguchi:1980jx} for details):

   \begin{description}
\item[Einstein] $C^{\pm}=0$ ($\Leftrightarrow R_{ab}=\tfrac{R}{4}g_{ab}$)
\item[Ricci flat]  $C^{\pm}=0,\quad s=0$ 
\item[Self-dual] $ \mathcal{A}_i=0 \Leftrightarrow \{W^-=0, \ C^{\pm}=0, \ s=0\}$  
\item[Anti-self-dual]  $ \mathcal{S}_i=0 \Leftrightarrow \{W^+=0, \ C^{\pm}=0, \ s=0\}$ 
\item[Conformally self-dual] $W^-=0$
\item[Conformally anti-self-dual] $W^+=0$
\item[Conformally flat]  $W^{+}=W^-=0$
\end{description}
Note that self-dual and anti-self-dual geometries are called half-flat in the mathematical literature, whereas self-dual and anti-self-dual is meant to be conformally self-dual and anti-self-dual.

\subsection{Einstein spaces}

 The self-dual and anti-self-dual geometries  have a special status as they are automatically Ricci flat:
 \begin{equation}\label{sdsec}
 \mathcal{A}_i=0 \ \mathrm{or} \  \mathcal{S}_i=0 \Rightarrow  C^{\pm}=0, \quad s=0.
 \end{equation}
They provide therefore special solutions of vacuum Einstein's equations, which include gravitational instantons already quoted in the introduction such as Eguchi--Hanson, Taub--NUT or Atiyah--Hitchin.

More general solutions are obtained by demanding conformal self-duality on Einstein spaces      
 \begin{equation}\label{csd-1}
W^+=0\ \mathrm{or} \ W^-=0 \quad \mathrm{and} \quad C^{\pm}=0
 \end{equation}
with non-vanishing scalar curvature\footnote{Requiring vanishing $C^\pm$ amounts to demanding the space to be Einstein ($R_{ab}=\frac{R}{4}g_{ab}$), which implies that its scalar curvature is constant.}     
\begin{equation}\label{csd-2}
s=2\Lambda. 
\end{equation}
Those are the \emph{quaternionic spaces} and  include other remarkable instantons such as Fubini--Study, Pedersen or Calderbank--Pedersen. 

Conditions (\ref{csd-1}) and (\ref{csd-2}) can be elegantly implemented by introducing the \emph{on-shell Weyl tensor}
\begin{equation}
\label{oswt}
 \widehat{\mathcal{W}}^{ab}=\mathcal{R}^{ab}-\frac{\Lambda}{3} \theta^a\wedge \theta^b.
\end{equation}
These 6 two-forms can be decomposed into self-dual and anti-self-dual parts:
\begin{eqnarray}
 \widehat{ \mathcal{W}}^+_i&=&\mathcal{S}_i-\frac{\Lambda}{6}\phi_i=\mathcal{W}_i^+ +\frac{1}{12}(s-2\Lambda)\phi_i+\frac{1}{2}C^+_{ij}\chi^j, \label{oswts}\\ 
 \widehat{ \mathcal{W}}^-_i&=&\mathcal{A}_i-\frac{\Lambda}{6}\chi_i=\mathcal{W}_i^- +\frac{1}{12}(s-2\Lambda)\chi_i+\frac{1}{2}C^-_{ij}\phi^j. \label{oswta}
\end{eqnarray}
Quaternionic spaces
are therefore obtained by demanding
\begin{equation}\label{csd-3}
 \widehat{ \mathcal{W}}^+_i=0\ \mathrm{or} \  \widehat{ \mathcal{W}}^-_i=0.
 \end{equation}
Furthermore, using the on-shell Weyl tensor (\ref{oswt}), the Einstein--Hilbert action reads:
 \begin{equation}
   S_{\mathrm{EH}}= \frac{1}{32\pi G}
   \int_{\mathcal{M}_4} \epsilon_{abcd}\left(\widehat{\mathcal{W}}^{ab} + \frac{\Lambda}{6} \theta^a\wedge \theta^b\right)\wedge \theta^c\wedge \theta^d.
  \end{equation}

\section{Self-dual gravitational instantons in Bianchi IX} \label{sdBIX}

Inspired by applications to homogeneous cosmology (see e.g. \cite{Ryan:1975jw}), spaces 
$\mathcal{M}_4$ topologically  equivalent to $ \mathbb{R}\times\mathcal{ M}_3$ have been investigated extensively in the cases where $\mathcal{ M}_3$ are homogeneous of Bianchi type. These foliations admit a three-dimensional group of motions acting transitively on the leaves $\mathcal{ M}_3$. 
 
The study of all Bianchi classes (I--IX) has been performed (for vanishing cosmological constant) in 
\cite{Kasner:1921zz, Lorenz:1983, Lorenz:1989} and more completed recently in  \cite{Bourliot:2009ad}. It turns out that only Bianchi IX exhibits a relationship  with quasimodular forms.

\subsection{Bianchi IX foliations}

Under the above assumptions, a metric on $\mathcal{M}_4$ can always be chosen as (see e.g. \cite{GBS}) 
\begin{equation}\label{met}
\mathrm{d}s^2= 
   \mathrm{d}t^2 
     + g_{ij}(t)  \sigma^i  \sigma^j, 
 \end{equation}
where $\sigma^i, i=1,2,3$ are the left-invariant Maurer--Cartan forms of the Bianchi group, satisfying
\begin{equation}
    \mathrm{d}\sigma^i = \frac{1}{2} c^i_{\hphantom{i}jk}\sigma^j \wedge \sigma^k.
 \end{equation}
This geometry admits three independent Killing vectors $\xi_i$, tangent to   $\mathcal{M}_3$ and such that
\begin{equation}
  \left[\xi_i, \xi_j\right] =c^i_{\hphantom{i}jk} \xi_k. 
 \end{equation}

In the case of Bianchi IX, the group is $SU(2)$. Using Euler angles,   the Maurer--Cartan forms read:
\begin{equation}
  \begin{cases}
\sigma^1= \sin\vartheta \sin\psi \, \mathrm{d}\varphi+\cos \psi \, \mathrm{d}\vartheta \\
\sigma^2= \sin\vartheta\cos\psi\, \mathrm{d}\varphi-\sin\psi\, \mathrm{d}\vartheta\\
\sigma^3=\cos\vartheta\, \mathrm{d}\varphi+\mathrm{d}\psi
\end{cases}
 \end{equation}
with $0\leq\vartheta\leq \pi, 0\leq\varphi\leq 2\pi, 0\leq\psi\leq {4\pi}$.
The structure constants are $c^i_{\hphantom{i}jk}=-\epsilon^i_{\hphantom{i}jk}=-\delta^{i\ell}\epsilon_{\ell jk}$ with $\epsilon_{123}=1$. Similarly the Killing vectors are
\begin{equation}
  \begin{cases}
  \label{LKil}
\xi_1= - \sin\varphi \cot\vartheta\, \partial_\varphi+\cos \varphi\, \partial_\vartheta+\frac{\sin \varphi}{\sin \vartheta} \, \partial_\psi \\
\xi_2=  \cos\varphi \cot\vartheta\, \partial_\varphi+\sin \varphi \,\partial_\vartheta-\frac{\cos \varphi}{\sin \vartheta} \, \partial_\psi \\
\xi_3= \partial_\varphi.
\end{cases}
\end{equation}

Although for some Bianchi groups it is necessary to keep $g_{ij}$ in (\ref{met}) general, for Bianchi IX it is always possible to bring it into a diagonal form, without loosing generality (for a systematic analysis of this, see \cite{Bourliot:2009ad}). We will make this assumption here, introduce three arbitrary functions of time $\Omega^i$ as well as a new time coordinate defined as  $\mathrm{d}t= \sqrt{ \Omega^1\Omega^2\Omega^3}\mathrm{d}T$, and write the most general metric (\ref{met}) on a Bianchi IX foliation as 
 \begin{equation}\label{metans}
\mathrm{d}s^2 =\delta_{ab}\, \theta^a\, \theta^b= \Omega^1\Omega^2\Omega^3\, 
\mathrm{d}T^2
    +
    \frac{\Omega^2\Omega^3}{\Omega^1}\left(\sigma^1\right)^2+
    \frac{\Omega^3\Omega^1}{\Omega^2}\left(\sigma^2\right)^2+
    \frac{\Omega^1\Omega^2}{\Omega^3}\left(\sigma^3\right)^2.
\end{equation}
For this metric, the two-form basis (\ref{2fs}) and (\ref{2fa}) reads:
\begin{eqnarray}
\phi^i&=&\Omega^j\Omega^k\mathrm{d}T\wedge\sigma^i +\Omega^i\sigma^j\wedge\sigma^k, \label{2fsIX}\\ 
\chi^i&=&\Omega^j\Omega^k \mathrm{d}T\wedge\sigma^i -\Omega^i\sigma^j\wedge\sigma^k,
\label{2faIX}\end{eqnarray}
where $i,j,k$ are a cyclic permutation of $1,2,3$ without over $i$.
Using Eqs. (\ref{torsles}), (\ref{metcon}) and (\ref{sdcon}), one finds for the corresponding Levi--Civita connection
 \begin{eqnarray}
     \Sigma_i&=&\frac{1}{4\sqrt{\Omega^1\Omega^2\Omega^3}}\left(
    \frac{\dot\Omega^i+\Omega^j\Omega^k}{\Omega^i}
    -\frac{\dot\Omega^j+\Omega^k\Omega^i}{\Omega^j}
    -\frac{\dot\Omega^k+\Omega^i\Omega^j}{\Omega^k}  
     \right) \theta^i\label{IX-sdcon},\\ 
    A_i&=&\frac{1}{4\sqrt{\Omega^1\Omega^2\Omega^3}}\left(
    \frac{\dot\Omega^i-\Omega^j\Omega^k}{\Omega^i}
    -\frac{\dot\Omega^j-\Omega^k\Omega^i}{\Omega^j}
    -\frac{\dot\Omega^k-\Omega^i\Omega^j}{\Omega^k}  
     \right) \theta^i,\label{IX-asdcon}
     \end{eqnarray}
where $\dot{f}$ stands for $\nicefrac{\mathrm{d}f}{\mathrm{d}T}$ (as previously, $i,j,k$ are a cyclic permutation of $1,2,3$ and no sum over $i$ is assumed) .

\subsection{First-order self-duality equations}\label{foe}

From now on, will focus  on self-dual solutions of Einstein vacuum equations (anti-self-dual solutions are related to the latter e.g. by time reversal). Following (\ref{sdcon}),  self-duality  
equations (\ref{sdsec}) read:
 \begin{equation}\label{sd-sec}
 \mathrm{d} A_i +\epsilon_{ijk} A^j \wedge A^k=0.
\end{equation}
Equations (\ref{sd-sec}) are second-order. They admit a first integral, algebraic in the anti-self-dual connection $A_i$:
 \begin{equation}\label{flsdLC}
A_i=\frac{\lambda_{ij}}{2}\sigma^j \quad \mathrm{with}\quad \lambda_{ij} =0 \ \mathrm{or}\ 
\delta_{ij}.
\end{equation}
Put differently, vanishing anti-self-dual Levi--Civita curvature can be realized either with a vanishing anti-self-dual connection, or with a specific non-vanishing one that can be set to zero upon appropriate local $SO(3)\subset SO(4)$ frame transformation (see \cite{Eguchi:1980jx} for a general discussion, \cite{Gibbons:1979xn} for Bianchi IX, or \cite{Bourliot:2009ad} for a more recent general Bianchi analysis). These two possibilities lead to two distinct sets of first-order equations. In the present case, using  (\ref{IX-asdcon}) one obtains:
 \begin{equation}\label{L}
A_i=0 \Leftrightarrow
\left\{\dot{\Omega}^1= \Omega^2 \Omega^3,\quad
   \dot{\Omega}^2= \Omega^3 \Omega^1,\quad
   \dot{\Omega}^3=\Omega^1 \Omega^2\right\},
\end{equation}
and
 \begin{equation}\label{DH}
A_i=\delta_{ij}\frac{\sigma^j}{2}
\Leftrightarrow\begin{cases}
    \dot{\Omega}^1 = \Omega^2 \Omega^3 - \Omega^1 \left(\Omega^2
      + \Omega^3 \right)   \\
    \dot{\Omega}^2 = \Omega^3 \Omega^1 - \Omega^2 \left(\Omega^3
      + \Omega^1 \right)   \\
    \dot{\Omega}^3 = \Omega^1 \Omega^2 - \Omega^3 \left(\Omega^1 +
      \Omega^2 \right).
  \end{cases}
\end{equation}

Historically, both systems were studied in the 19th century in the search of integrals lines of vector fields. The first is the \emph{Lagrange} system, appearing as an extension of the rigid-body equations of motion. It is algebraically integrable and was solved \emph{\`a la} Jacobi.  The second set is called \emph{Darboux--Halphen} and appeared in Darboux's work on triply orthogonal surfaces \cite{Darboux}. Generically, it does not possess any polynomial first integral, and was solved by Halphen in full generality using Jacobi theta functions \cite{halph}. 

In the late seventies, integrable systems of equations such as Lagrange or Darboux--Halphen emerged in a systematic manner in self-dual Yang--Mills reductions \cite{Ward85}. This has led many authors to investigate these equations in great detail and, in particular, to unravel their rich integrability properties (see e.g. \cite{Takhtajan:1992qb, maciejewski95, Ablowitz:2003bv} as a sample of the dedicated literature). It took a long time, however, to realize that these systems were actually related with gravitational instantons, foliated by squashed spheres. 

When all three $\Omega^i$s are identical, the leaves of the foliation are isotropic three-spheres with $SU(2)\times SU(2)$ isometry generated by the above left Killing vectors $\xi_i, i=1,2,3$ (\ref{LKil}), as well as by three right Killing vectors 
\begin{equation}
  \begin{cases}
e_1= - \sin\psi \cot\vartheta\, \partial_\psi+\cos \psi\, \partial_\vartheta+\frac{\sin \psi}{\sin \vartheta} \, \partial_\varphi \\
e_2= - \cos\psi \cot\vartheta\, \partial_\psi-\sin \psi \,\partial_\vartheta+\frac{\cos \psi}{\sin \vartheta} \, \partial_\varphi \\
e_3= \partial_\psi.
\end{cases}
\end{equation}
Lagrange and Darboux--Halphen  systems are equivalent in this case (actually related by time reversal), $\dot{\Omega}=\pm \Omega^2$, and the solutions lead to flat Euclidean four-dimensional space. 

More interesting is the case where $\Omega^1=\Omega^2\neq \Omega^3$. Here, the leaves are axisymmetric  squashed three-spheres,  invariant under  an $SU(2)\times U(1)$ isometry group generated by $\xi_i, i=1,2,3$ and $e_3$. On the one hand, the Lagrange system leads to two distinct gravitational instantons known as Eguchi--Hanson I and II \cite{Eguchi:1978gw, Eguchi:1979yx}, out of which the first has  a naked singularity and is usually discarded. On the other hand, the Darboux--Halphen equations deliver the celebrated Taub--NUT instanton \cite{Taub-nut}.

Thanks to the algebraic integrability properties of Lagrange system, it took only a few months to Belinski \emph{et al.} to generalize the Eguchi--Hanson solution to the case where $\Omega^1\neq \Omega^2\neq \Omega^3$ \cite{Belinsky:1978ue} -- the symmetry is strictly $SU(2)$ but the solution is plagued with naked singularities. A similar generalization of the Taub--NUT solution turned out much more intricate, and after some fruitless attempts \cite{Gibbons:1979xn}, Atiyah and Hitchin reached a regular solution, eligible as a gravitational instanton and  expressed in terms of elliptic functions \cite{Atiyah1}. It was only realized in 1992 by Takhtajan \cite{Takhtajan:1992qb} that first-order self-duality equations for Bianchi IX gravitational instantons were in fact Lagrange and  Darboux--Halphen systems, and that the Atiyah--Hitchin instanton was a particular case of the general solution found by Halphen in 1881 \cite{halph}.

It is finally worth mentioning that the above systems of ordinary differential equations also appear in the framework of geometric flows in three-dimensional Bianchi IX homogeneous spaces. The original mention on that matter can be found in \cite{Cvetic:2001zx}; later and independently it was also quoted in \cite{Sfetsos:2006}. At that original stage, this relationship was limited to the case of Bianchi IX with diagonal metric. It was proven recently to hold in full generality in all Bianchi classes \cite{Bourliot:2009fr,Petropoulos:2011qq}. 

\section{The Darboux--Halphen system}

The Darboux--Halphen branch of the self-duality first-order equations of Bianchi IX foliations in vacuum is the most interesting for our present purpose as it is the one related with quasimodular forms.  
 
 \subsection{Solutions and action of $SL(2, \mathbb{C} )$}
 
Consider the system in the complex plane: $\omega^i(z)$, $z\in \mathbb{C}$ satisfying 
\begin{equation}
\begin{cases}
\frac {\mathrm{d}\omega^1}{\mathrm{d}z} = \omega^2 \omega^3 - \omega^1 \left(\omega^2
      + \omega^3 \right)  \\
   \frac {\mathrm{d}\omega^2}{\mathrm{d}z} = \omega^3 \omega^1 - \omega^2 \left(\omega^3
      + \omega^1 \right)   \\
  \frac {\mathrm{d}\omega^3}{\mathrm{d}z}  = \omega^1 \omega^2 - \omega^3 \left(\omega^1 +
      \omega^2 \right).
\end{cases}
\end{equation}
The general solutions of this system have the following properties \cite{halph, Takhtajan:1992qb}:
 \begin{itemize}
    \item The $\omega$s are regular, univalued and holomorphic in a
    region with movable boundary (\emph{i.e.} a dense set of essential
    singularities). The location of this boundary accurately determines the
    solution.
    \item If $\omega^i(z)$ is a solution, thus
\begin{equation}\label{slact}
 \tilde \omega^i (z) = \frac{1}{\left(c z + d  \right)^2} \omega^i
    \left( \frac{a z + b}{c z + d} \right) + \frac{c}{c z +
    d}, \quad  \left(\begin{matrix}
      a & b \\ c & d
    \end{matrix}
  \right) \in SL(2, \mathbb{C} )
\end{equation}
is another solution\footnote{The same property holds for the Lagrange system (\ref{L}),
limited to the subgroup of transformations of the form $\left(\begin{smallmatrix}
      a & b \\ 0 & \nicefrac{1}{a}
    \end{smallmatrix}\right)$. This solution-generating pattern based on the $SL(2,\mathbb{R})$ is closely related to the Geroch method \cite{Geroch:1970nt, Bossard:2012u}.} with singularity  boundary moved according to $z\to\frac{a z + b}{c z + d}$. 
\end{itemize}

The resolution of the equations and the nature of the solutions strongly depend on whether the $\omega$s are different or not. In the case where $\omega^1=\omega^2=\omega^3$, the solution is simply
\begin{equation}\label{alleq}
\omega^{1,2,3}= \frac{1}{z-z_0}
\end{equation}
with $z_0$ an arbitrary constant. Under  $SL(2,\mathbb{C})$,  the new solution $\tilde \omega^i (z) $ is of the form (\ref{alleq}) with the pole displaced according  to
\begin{equation}
\tilde z_0= -\frac{d z_0-b}{c z_0-a}.
\end{equation}
 If    $\omega^1=\omega^2\neq\omega^3 $ the solutions are still  algebraic:
\begin{equation}\label{2eq}
\omega^{1,2}= \frac{1}{z-z_0}, \quad \omega^3=
\frac{z-z_\ast}{(z-z_0)^2}
\end{equation}
with two arbitrary constants: $z_0, z_\ast$. A simple pole for $\omega^{1,2}$, and double for
$\omega^3$ appears at  $z_0$, whereas $ z_\ast$ is a root for  $\omega^3$. Acting with $SL(2,\mathbb{C})$ keeps the structure (\ref{2eq}) with new parameters: 
\begin{equation}
 \tilde z_0= -\frac{d z_0-b}{c z_0-a}, \quad \tilde z_0- \tilde z_\ast = \frac{z_0- z_\ast }{\left(c z_0-a\right)^2}.
\end{equation}

The fully anisotropic case is our main motivation here. In this case
no  algebraic first integrals exist and the general solution (see
\cite{halph, Takhtajan:1992qb, maciejewski95}) is expressed in terms
of quasimodular forms,  $\omega^i \in
QM^1_2\left(\Gamma(2)\right)$, where $\Gamma(2)$ is the level-2 congruence subgroup of $SL(2,\mathbb{Z})$ (the subset of elements of the form $\left(\begin{smallmatrix}
      a & b \\ c & d
    \end{smallmatrix}\right)
  =\left( \begin{smallmatrix}
      1 & 0 \\ 0 & 1
    \end{smallmatrix} \right) \ \mathrm{mod} \ 2 $) . Concretely
\begin{equation}\label{DH-sol}
 \omega^i (z)=
    - \frac{1}{2} \frac{\mathrm{d }}{\mathrm{d }z } \log \mathcal{E}^i (z) 
\end{equation}
with $\mathcal{E}^i (z)$ triplet\footnote{\label{modtran}Notice their general transformations as generated by $ z\to-\nicefrac{1}{z}$ and $ z+1$ :
\begin{eqnarray}
 z\to-\nicefrac{1}{z}&:&\quad \begin{pmatrix}
   \mathcal{E}_1&
   \mathcal{E}_2&
   \mathcal{E}_3
  \end{pmatrix} \to  z^2   \begin{pmatrix}
   \mathcal{E}_2 &
   \mathcal{E}_1 &
   -\mathcal{E}_3
  \end{pmatrix}\nonumber\\
   z\to z+1&:&\quad \begin{pmatrix}
   \mathcal{E}_1&
   \mathcal{E}_2&
   \mathcal{E}_3
  \end{pmatrix} \to  - \begin{pmatrix}
   \mathcal{E}_3 &
   \mathcal{E}_2 &
   \mathcal{E}_1
  \end{pmatrix}.\nonumber
 \end{eqnarray}}  of  \emph{holomorphic weight-2 modular forms} of  $\Gamma(2)$.
Again,  the $SL(2,\mathbb{C})$ action (\ref{slact}) generates new solutions $\left\{\omega^i\right\}\to \left\{\tilde\omega^i\right\}$ with a displaced set of singularities in $\mathbb{C}$, whereas the $SL(2,\mathbb{Z})\subset SL(2,\mathbb{C})$ acts as a permutation on $\omega$s.

Note for completeness that real solutions of the real coordinate $T$ are obtained from the general solutions as 
\begin{equation}
\Omega^\ell (T) = i  \omega^\ell(iT) = - \frac{1}{2}
\frac{\mathrm{d } }{\mathrm{d } T } \log \mathcal{E}^\ell (iT).
\end{equation}
According to (\ref{slact}), new real solutions are generated as 
\begin{equation}\label{slactR}
 \tilde \Omega^i (T) = \frac{1}{\left(C T + D  \right)^2} \Omega^i
    \left( \frac{A T + B}{C T + D} \right) + \frac{C}{C T +
    D}, \quad  \left(\begin{matrix}
      A & B \\ C & D
    \end{matrix}
  \right) \in SL(2, \mathbb{R}).
\end{equation}

\subsection{Relationship with Schwartz's and Chazy's equations}
   
Anisotropic solutions of  the Darboux--Halphen system  ($\omega^1\neq \omega^2\neq\omega^3$) exhibit relationships with other remarkable equations.
 Define
\begin{equation}
 \lambda=\frac{\omega^1-\omega^3}{\omega^1-\omega^2}.
 \end{equation}
For $\omega^i$ solving Darboux--Halphen equations, $\lambda$ is a solution of  
of Schwartz's equation
\begin{equation}\label{sch}
    \frac{\lambda'''}{\lambda'}-\frac{3}{2}\left(\frac{\lambda''}{\lambda'}\right)^2=-\frac{1}{2}\left(
    \frac{1}{\lambda^2}+\frac{1}{(\lambda-1)^2}-\frac{1}{\lambda(\lambda-1)}
        \right)
\left(\lambda'\right)^2 .  
 \end{equation}
Conversely, any solution of the latter equation provides a solution for the Darboux--Halphen system as the following triplet:
\begin{equation}\label{schsol}
     \mathcal{E}^1 = \frac{\nicefrac{\mathrm{d } \lambda}{\mathrm{d } z}}{\lambda},
  \quad \mathcal{E}^2 = \frac{\nicefrac{\mathrm{d } \lambda}{\mathrm{d } z}}{\lambda -1},
   \quad \mathcal{E}^3 =\frac{\nicefrac{\mathrm{d } \lambda}{\mathrm{d } z}}{\lambda(\lambda
   -1)},
 \end{equation}
 from which it is straightforward to show that
 \begin{equation}
  \mathcal{E}^1-\mathcal{E}^2+\mathcal{E}^3=0.
 \end{equation}
We also quote for completeness 
    \begin{equation}
\frac{1}{1-\lambda}  =  \frac{\omega^1-\omega^2}{\omega^3-\omega^2},\quad 
\frac{1-\lambda}{\lambda} =    \frac{\omega^3-\omega^2}{\omega^1-\omega^3}.
    \end{equation}

Define now
    \begin{equation}
   y=-2\left(\omega^1+\omega^2+\omega^3\right).
    \end{equation}
Again, for  solutions of  Darboux--Halphen equations $\omega^i$,
$y$ a solution of Chazy's equation \cite{Chazy}
    \begin{equation} 
y'''=2yy'' - 3(y')^2.
    \end{equation}
The first and second derivatives of $y$ provide the remaining symmetric products 
\begin{eqnarray}
y'&=&2\left(\omega^1\omega^2+\omega^2\omega^3+\omega^3\omega^1\right),\\
y''&=&-12 \omega^1\omega^2\omega^3.
 \end{eqnarray}
The Jacobian relating  $\{\omega^1,\omega^2,\omega^3\}$ to $ \{y,y',y''\}$,
    \begin{equation} 
 J=(\omega^1-\omega^2)(\omega^2-\omega^3)(\omega^3-\omega^1)
    \end{equation}
 is regular for $\omega^1\neq\omega^2\neq\omega^3$. The latter are alternatively obtained by solving the cubic equation
    \begin{equation} 
\omega^3+\frac{1}{2}y\omega^2 +\frac{1}{2}y'\omega+\frac{1}{12}y'' =0,
    \end{equation}
 for any solution $y$ of Chazy's equation.

   \subsection{The original Halphen solution}
 A particular solution of the Darboux system (\ref{DH}) is 
the original Halphen solution \cite{halph}. In this language,  it corresponds to
   $\lambda_{\mathrm{H}} = \nicefrac{\vartheta_2^4}{\vartheta_3^4}$: 
    \begin{equation}\label{halphsol} 
\begin{cases}
     \mathcal{E}^1_{\mathrm{H}} = i\pi\vartheta_4^4 \\
     \mathcal{E}^2_{\mathrm{H}} = -i\pi \vartheta_2^4  \\
     \mathcal{E}^3_{\mathrm{H}} = -i\pi\vartheta_3^4
  \end{cases}
\Leftrightarrow
\begin{cases}
    \omega^1_{\mathrm{H}} = \frac{\pi}{6i}\left(E_2-\vartheta_2^4 - \vartheta_3^4 \right) \\
    \omega^2_{\mathrm{H}} = \frac{\pi}{6i}\left(E_2+\vartheta_3^4 + \vartheta_4^4 \right)  \\
    \omega^3_{\mathrm{H}} = \frac{\pi}{6i}\left(E_2+\vartheta_2^4 - \vartheta_4^4
    \right).
  \end{cases}
    \end{equation}
This is also  the solution found by Atiyah and Hitchin \cite{Atiyah1} as the
Bianchi IX gravitational instanton solution relevant for describing the configuration
space of two slowly moving BPS $SU(2)$ Yang--Mills--Higgs 
monopoles \cite{Manton:1981mp,Gibbons:1986df}.
The corresponding Chazy's solution and derivatives are combinations of (holomorphic)
Eisenstein series (see appendix, 
Eqs.  \eqref{A6}):
   \begin{equation} 
\begin{cases}
   y_{\mathrm{H}}=i\pi E_2 \\
   y_{\mathrm{H}}'=\frac{i\pi}{6} \left(E_2^2-E_4\right)   \\
   y_{\mathrm{H}}''=-\frac{i\pi^3}{18} \left(E_2^3-3E_2E_4+2E_6\right).   
    \end{cases}
    \end{equation}
Starting from (\ref{halphsol}) all solutions are obtained by $SL(2,\mathbb{C})$ action (\ref{slact}).

   \section{Back to Bianchi IX self-dual solutions}
   
   Any real solution $\{\Omega^i(T)\}$ of the Darboux--Halphen system provides a four-dimensional self-dual solution of Einstein vacuum equations in the form (\ref{metans}). Not all these solutions are however \emph{bona fide} gravitational instantons as some regularity requirements must be fulfilled.
   
     \subsection{Some general properties}
   
An elementary consistency requirement is that the metric (\ref{metans})  should not change sign along
$T$. In particular, a simple root of a single $\Omega^i$ turns out to be a genuine curvature singularity. Assuming e.g. linearly vanishing $\Omega^1$ and introducing as time coordinate the proper time $\tau$ around the root, the metric locally reads:
    \begin{equation} \label{singmet}
 \mathrm{d}s^2 \approx  \mathrm{d}\tau^2 +
 \frac{\Xi}{\tau^{\nicefrac{2}{3}}}\left(\sigma^1\right)^2
 + \Upsilon
 \tau^{\nicefrac{2}{3}} \left(\left(\sigma^2\right)^2+\left(\sigma^3\right)^2\right)
    \end{equation}
with $\Xi, \Upsilon$ constants.
This metric has a curvature singularity at $\tau=0$.
  
Other pathologies can appear, which do not necessarily affect the consistency of the solution. Poles of some $\Omega$s or multiple roots are potential natural boundaries or (non-)removable coordinate singularities such as \emph{bolts or nuts}. The latter are fixed points of some Killing vectors $\xi$ ($\nabla_{(\nu}\xi_{\mu)}=0$), for which the matrix $\nabla_{[\nu}\xi_{\mu]}$ is respectively of rank 2 and 4. A general, complete and detailed presentation of these properties is beyond the scope of these notes and is available in the original paper \cite{nuts}. For our purpose here, we recall two generic situations, where again we present the metric in local proper time $\tau$ around a fixed point at $\tau=0$:
\begin{description}
\item[Rank 2 --  bolt]  This singularity is removable if the metric behaves as
    \begin{equation}  \label{gebolt}
\begin{array}{rcl}
 \mathrm{d}s^2 &\approx&  \mathrm{d}\tau^2
+ \zeta^2
\left(\left(\sigma^1\right)^2+\left(\sigma^2\right)^2\right) +\frac{n^2\tau^2}{4} \left(\sigma^3\right)^2\\ &=&
 \mathrm{d}\tau^2
+ \frac{n^2\tau^2}{4} \left(
 \mathrm{d}\psi + \cos \vartheta  \mathrm{d}\varphi
\right)^2+ \zeta^2 \left( \mathrm{d}\vartheta^2 +
\sin^2\vartheta \mathrm{d}\varphi^2
 \right),
\end{array}
    \end{equation}
and provided $\nicefrac{n\psi}{2}\in[0,2\pi[$.  Locally the geometry is thus $\mathbb{R}^2\times S^2$.
\item[Rank 4 -- nut]  This singularity is removable if  the metric behaves as
    \begin{equation} \label{gennut}
\begin{array}{rcl}
 \mathrm{d}s^2&\approx& \mathrm{d}\tau^2
+\frac{\tau^2}{4} 
\left(\left(\sigma^1\right)^2+\left(\sigma^2\right)^2+\left(\sigma^3\right)^2\right) \\ &=& \mathrm{d}\tau^2 + \frac{\tau^2}{4} \left(
\mathrm{d}\vartheta^2 + \sin^2\vartheta \mathrm{d}\varphi^2
+\left(
 \mathrm{d}\psi + \cos \vartheta  \mathrm{d}\varphi
\right)^2
 \right).
 \end{array}
    \end{equation}
For a nut, the local geometry is $\mathbb{R}^4$ (here in polar coordinates).
\end{description}
 
\subsection{Behaviour of  Darboux--Halphen solutions} \label{geombDH}
 
\begin{figure}[!h]
\begin{center}
\includegraphics[height=8cm]{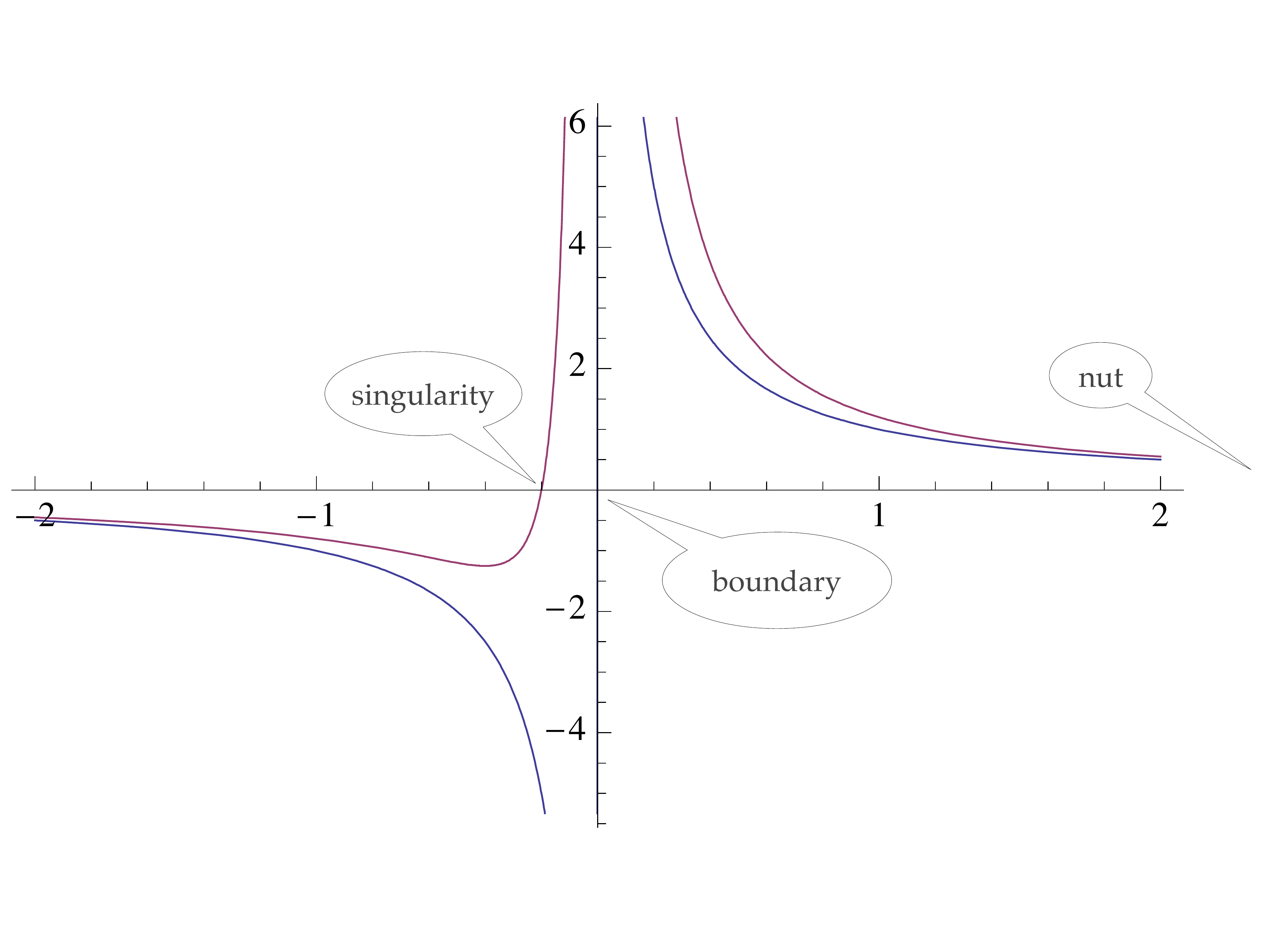}
\end{center}
\caption{Generic solution $\Omega^1=\Omega^2<\Omega^3$.}\label{TN}
\end{figure}

As already pointed out, there are thee distinct cases to consider: $\Omega^i$ all equal,  $\Omega^1=\Omega^2\neq \Omega^3$ or  $\Omega^1\neq\Omega^2\neq \Omega^3$. In the first case, 
\begin{equation} \label{alleq-R}
\Omega^i=\frac{1}{T-T_0} \quad \forall i,
\end{equation}
 and the four-dimensional solution corresponds to flat space. When only two $\Omega$s  are equal, the isometry group is extended to $SU(2)\times U(1)$ and real solutions read:
    \begin{equation} \label{2eq-R}
\Omega^{1,2}= \frac{1}{T-T_0}, \quad \Omega^3=
\frac{T-T_\ast}{(T-T_0)^2}.
\end{equation}
 There are 3 special points: $T= T_\ast, T_0$ and 
  $T\to\infty$. One can analyze their nature by zooming around them, using proper time:
  \begin{itemize}
\item At $T\to \infty $ one recovers  the behaviour (\ref{gennut}) and this point is a nut.
\item Around $T=T_0$ one finds
    \begin{equation} 
 \mathrm{d}s^2\approx\mathrm{d}\tau^2 + \tau^2
\left( \mathrm{d}\vartheta^2 + \sin^2\vartheta \mathrm{d}\varphi^2
\right) +
\frac{1}{T_0-T_\ast}
\left(
 \mathrm{d}\psi + \cos \vartheta  \mathrm{d}\varphi
\right)^2.
\end{equation}
This is  an $S^1$ fibration over $\mathbb{R}^3$, the  fiber being $\mathrm{d}\psi + \cos \vartheta  \mathrm{d}\varphi$. It is  called Taubian infinity (see \cite{nuts}) and appears as a natural ``boundary''.
\item At $T =  T_\ast$ there is a curvature singularity as the metric behaves like (\ref{singmet})
with $\Xi=\left(\frac{2}{3(T_0-T_\ast)^2}\right)^{\nicefrac{2}{3}}$, $\Upsilon=\left(\frac{3}{2(T_0-T_\ast)}\right)^{\nicefrac{2}{3}}$.
\end{itemize}
One therefore concludes that in order to avoid the presence of naked singularities, the singular point $T=T_\ast$ should be hidden behind the Taubian infinity \emph{i.e.} $T_\ast<T_0$ (see Fig. \ref{TN}). Under this assumption, the self-dual solution at hand is well behaved and provides the Taub--NUT gravitational instanton \cite{Taub-nut, Eguchi:1980jx}. It is most commonly written as:
 \begin{equation} 
\mathrm{d}s^2=
\frac{r+m}{r-m}\frac{\mathrm{d}r^2}{4}
+\frac{1}{4}\left(r^2-m^2\right)\left(\left(\sigma^1\right)^2+
\left(\sigma^2\right)^2\right)+m^2\frac{r-m}{r+m}\left(\sigma^3\right)^2,
\end{equation}
where 
$m^2 =\frac{1}{T_0-T_\ast}>0$ and $m(r-m)=\frac{2}{T-T_0}$.

\begin{figure}[!h]
\begin{center}
\includegraphics[height=8cm]{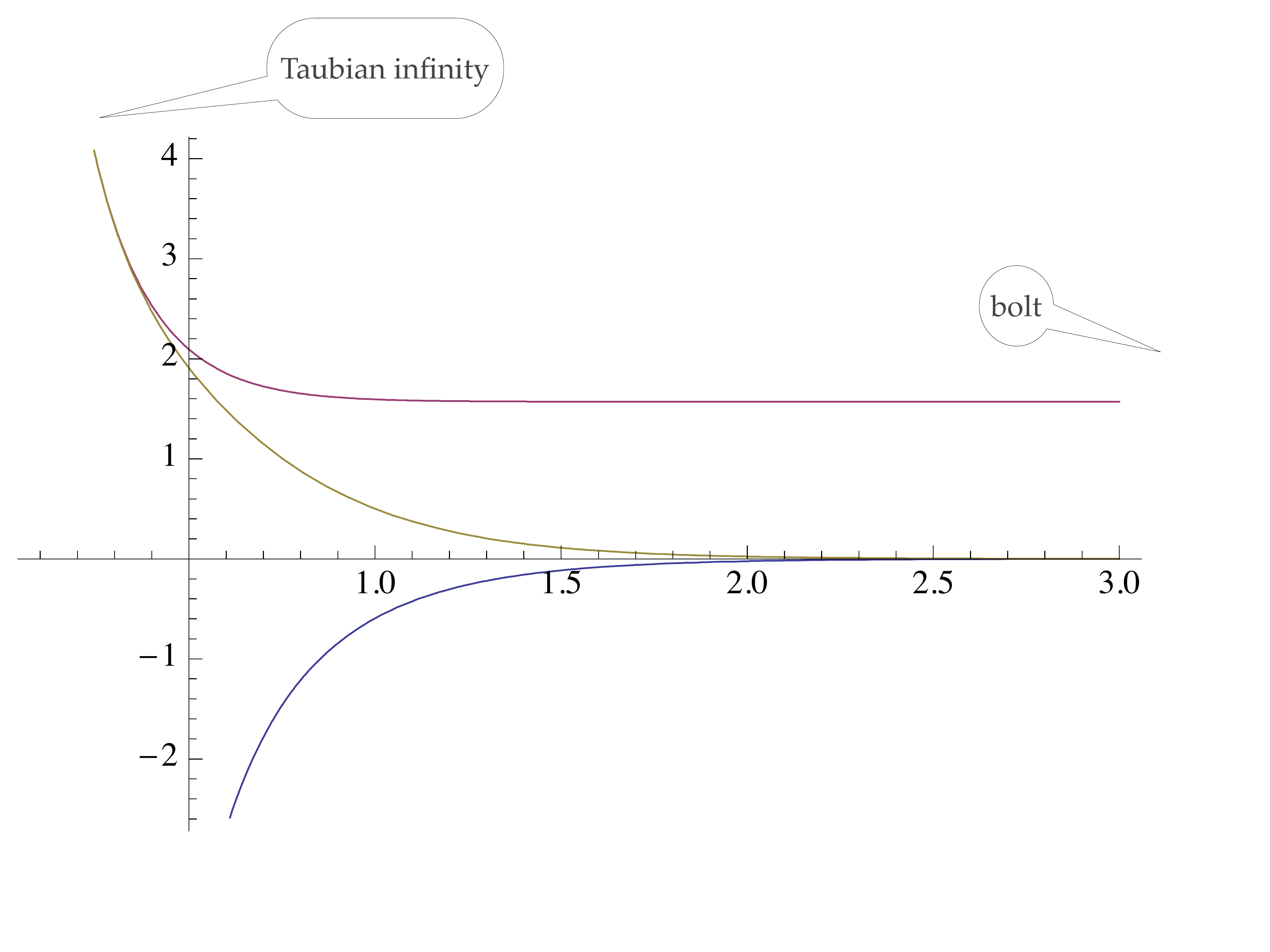}
\end{center}
\caption{Halphen original solution ($\Omega^1_{\mathrm{H}}<0<\Omega^3_{\mathrm{H}}<\Omega^2_{\mathrm{H}}$).} \label{halphorig}
\end{figure}

The case where  $\Omega^1\neq\Omega^2\neq \Omega^3$ is the most interesting in the present context since it involves quasimodular forms. The real Halphen solution (see Eq. (\ref{halphsol}) with $z=iT$ or $q=\exp -2\pi T$)) reads:
 \begin{equation} \label{halphsolR}
\begin{cases}
    \Omega^1_{\mathrm{H}}(T) = \frac{\pi}{6}\left(E_2-\vartheta_2^4 - \vartheta_3^4 \right) <0\\
    \Omega^2_{\mathrm{H}}(T) = \frac{\pi}{6}\left(E_2+\vartheta_3^4 + \vartheta_4^4 \right)  \\
    \Omega^3_{\mathrm{H}}(T)= \frac{\pi}{6}\left(E_2+\vartheta_2^4 - \vartheta_4^4
    \right)< \Omega^2_{\mathrm{H}} .
  \end{cases}
\end{equation}
It is defined for $T>0$ with a pole at $T=0$:
 \begin{equation} 
\Omega^1_{\mathrm{H}}\approx -\frac{\pi}{2T^2},\quad
    \Omega^{2,3}_{\mathrm{H}}\approx \frac{1}{T}.
\end{equation}
Around this pole, the behaviour of the metric is 
 \begin{equation} 
 \mathrm{d}s^2 \approx -\left(  \mathrm{d}\tau^2
+\tau^2 \left(\left(\sigma^3\right)^2+\left(\sigma^2\right)^2\right)+ \frac{2}{\pi}
\left(\sigma^1\right)^2\right),
\end{equation}
and we recover a Taubian infinity ($S^1$ fiber over $\mathbb{R}^3$). The large-$T$ behaviour is exponential towards a constant 
 \begin{equation}   \label{lTHo}   
      \Omega^{1,3}_{\mathrm{H}}\approx \mp 4\pi \exp{-\pi T},\quad
    \Omega^{2}_{\mathrm{H}}\approx \nicefrac{\pi}{2}+ 4\pi \exp{-2\pi T}
\end{equation}
with
 \begin{equation}   
   \mathrm{d}s^2 \approx-\left(  \mathrm{d}\tau^2
+ \frac{\pi}{2}
\left(\left(\sigma^1\right)^2+\left(\sigma^3\right)^2\right) +4\tau^2 \left(\sigma^2\right)^2\right).
\end{equation}
This is precisely a bolt as in Eq. (\ref{gebolt}) with $n=4, \zeta=\sqrt{\nicefrac{\pi}{2}}$ and permutation of principal directions 2 and 3.
All this is depicted in Fig. \ref{halphorig}. 

As already quoted, the self-dual vacuum geometry corresponding to Halphen's original solution is  the Atiyah--Hitchin  gravitational instanton \cite{Atiyah1}.  Using modular transformations (\ref{slactR}), one constructs all other real solutions with strict $SU(2)$ isometry (\emph{i.e.} with all $\Omega^i$ different):
 \begin{equation}  \label{slHactR}    
\Omega^i (T) = \frac{1}{\left(CT + D \right)^2} \Omega^i_{\mathrm{H}}
    \left( \frac{A T+ B}{CT + D} \right) + \frac{C}{CT +
    D}.
\end{equation}
 Are those well behaved? 
      
The answer is no because a root of one $\Omega$ always appears between the Taubian infinity and the bolt. This root is a curvature singularity, which spoils the regularity of the solution.  In order to elaborate on that, we first observe that in Eq. (\ref{slHactR}), $\frac{A T+ B}{CT + D}$ must be positive, as real $ \Omega^i_{\mathrm{H}}$ are only defined for positive argument. Assume for concreteness that
 \begin{equation}     
    \lim_{T\to \infty}\frac{A T+ B}{CT +
    D}=\frac{A}{C}>0. 
\end{equation}
 On the one hand, at large $T$
 \begin{equation}     
\Omega^i = \frac{1}{T} + \mathrm{O}\left(\nicefrac{1}{T^2}\right),
\end{equation}
and trading $T$ for the local proper time one finds a nut (see (\ref{gennut})).  On the other hand, the values
$T_\infty=-\nicefrac{D}{C}<T_0=-\nicefrac{B}{A}$ correspond to two poles, and 
$\Omega^i (T)$ are defined for $T<T_\infty$  or $T_0<T$ with reflected behaviour.
  For $T_0 \lessapprox T$
 \begin{equation}    
     \Omega^1 \approx
    -\frac{\pi}{2A^2}\frac{1}{\left(T-T_0\right)^2}, \quad \Omega^{2,3} \approx \frac{1}{T-T_0}
\end{equation}
(note the sign flip in $\Omega^1$) and
 \begin{equation}  
        - \mathrm{d}s^2\approx\mathrm{d}\tau^2 + \tau^2 \left(
\left(\sigma^2\right)^2+\left(\sigma^3\right)^2\right) + \frac{2A^2}{\pi}\left(\sigma^1\right)^2.
\end{equation}
Therefore $T=T_0$ is a Taubian infinity ($S^1$ fiber over $\mathbb{R}^3$), and as $T$ moves from 
$T=T_0$ to $T\to +\infty$ one moves from the Taubian infinity (``boundary'') to a nut. A similar conclusion is reached when scanning $T$ from $T_\infty$ to $ -\infty$. 

\begin{figure}[!h]
\begin{center}
\includegraphics[height=8.cm]{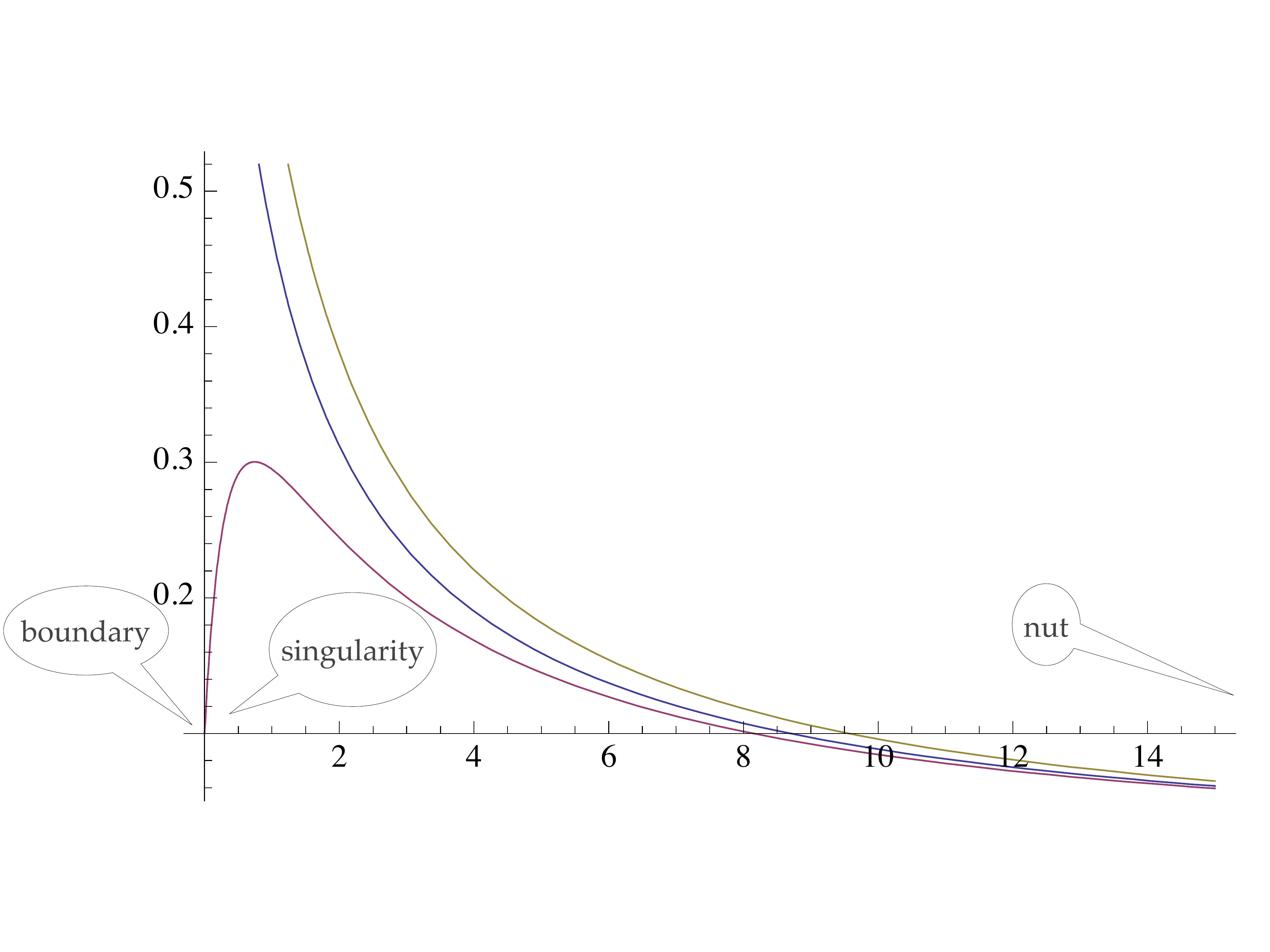}
\end{center}
\caption{Generic solution for $T_0<T$ and
$0<\Omega^1<\Omega^3<\Omega^2$.}  \label{halphanis}
\end{figure}

The problem arises because there is always a value $T_\ast$ such that  $T_0<T_\ast<\infty$ with $\Omega^1_\ast =    0<\Omega^3_\ast<\Omega^2_\ast$ (see Fig. \ref{halphanis}). This unavoidable root is a genuine curvature singularity of the metric. Because of this, no anisotropic solution of the Darboux--Halphen system  other than the original one ((\ref{halphsol}) or (\ref{halphsolR})) provides a well-behaved Bianchi IX gravitational instanton.
  
 \subsection{A parenthesis on Ricci flows}

Ricci flows describe the evolution of a metric on a manifold, governed by the following first-order equation:
\begin{equation}  \label{rf}
\frac{\partial g_{ij}}{\partial t}=-R_{ij},
\end{equation}
where $R_{ij}$ stands for the Ricci tensor of the Levi--Civita connection associated with $g_{ij}$ (see e.g. \cite{chow:2004}).
It was introduced by Hamilton in 1981 \cite{Hamilton82} in order to gain insight into the geometrization conjecture of Thurston (see e.g. \cite{Thurston78}), a generalization of  Poincar\'e's 1904 conjecture for three-manifolds, finally demonstrated by  Perel'man in 2003 \cite{perel:2003}. Ricci flows are also important in modern physics as they describe the renormalization group evolution in two-dimensional sigma-models \cite{Friedan:1980jm}.

The case of homogeneous three-manifolds is important as it appears in the final stage of 
Thurston's geometrization. Homogeneous three-manifolds include all 9
Bianchi groups plus 3 coset spaces, which are $H_3$, $H_2\times S^1$,
$S^2\times S^1$ ($S^n$ and $H_n$ are spheres and hyperbolic spaces respectively) \cite{Milnor,  Scott}. The general asymptotic behaviour was studied in detail in \cite{Isenberg:1992}. A remarkable and already quoted result \cite{Cvetic:2001zx, Sfetsos:2006, Bourliot:2009fr, Petropoulos:2011qq} is the relationship between 
the parametric evolution of a metric
\begin{equation}     
 \mathrm{d}\tilde{s}^2 =
   \sqrt{ \frac{\Omega^2\Omega^3}{\Omega^1}}\left(\sigma^1\right)^2+
    \sqrt{  \frac{\Omega^3\Omega^1}{\Omega^2}}\left(\sigma^2\right)^2+
     \sqrt{\frac{\Omega^1\Omega^2}{\Omega^3}}\left(\sigma^3\right)^2
     \end{equation}
on $\mathcal{M}_3$ of Bianchi type\footnote{The precise statement is actually formulated for more general, non-diagonal metrics, as explained in detail in \cite{Petropoulos:2011qq}, and is valid in all Bianchi classes. For Bianchi IX, the diagonal ansatz exhausts, however, all possibilities.},
 and the time evolution inside a self-dual gravitational instanton on $\mathcal{M}_4=\mathbb{R}\times\mathcal{M}_3$ as given in (\ref{metans}): the equations are the same ($t$ in (\ref{rf}) and $T$ in (\ref{metans}) are related as $\mathrm{d}t=\sqrt{\Omega^1\Omega^2\Omega^3}
\mathrm{d}T$). Ricci flow on three-spheres is therefore 
governed by the Darboux--Halphen equations (\ref{DH}).
 
Solutions  of  the Darboux--Halphen system describe Ricci-flow evolution if $\forall i \ \Omega^i(T) > 0 $, assuming that this holds at some initial time $T_0$. It is straightforward to see that this is always guaranteed. Indeed, it is true when at least two $\Omega$s are equal, as one can see directly from the algebraic solutions (\ref{alleq-R}) and (\ref{2eq-R}). More generally,
suppose that $0<\Omega^1_0<\Omega^2_0<\Omega^3_0$ (the subscript refers to the initial time $T_0$) and that
$\Omega_1$ has reached at time $T_1$ the value $\Omega_1^1=0$,
while $\Omega^2_1, \Omega^3_1>0$. From Eqs.~(\ref{DH}) we
conclude that at time $T_1$,
$\dot{\Omega}_2^1=\dot{\Omega}_3^1=-\Omega_2^1\, \Omega_3^1<0$ and
$\dot{\Omega}_1^1=\Omega_2^1\, \Omega_3^1>0$. This latter
inequality implies that $\Omega_1$ vanishes at $T_1$ while it is
increasing, passing therefore from negative to positive values.
This could only happen if $\Omega^1_0$ were negative, which
contradicts the original assumption. However, if indeed
$\Omega^1_0<0$ and $\Omega^2_0,\Omega^3_0>0$, there is a time
$T_1$ where $\Omega^1$ becomes positive and remains positive
together with $\Omega^2$ and $\Omega^3$ until they reach the
asymptotic region.

\begin{figure}[!h]
\begin{center}
\includegraphics[height=8.cm]{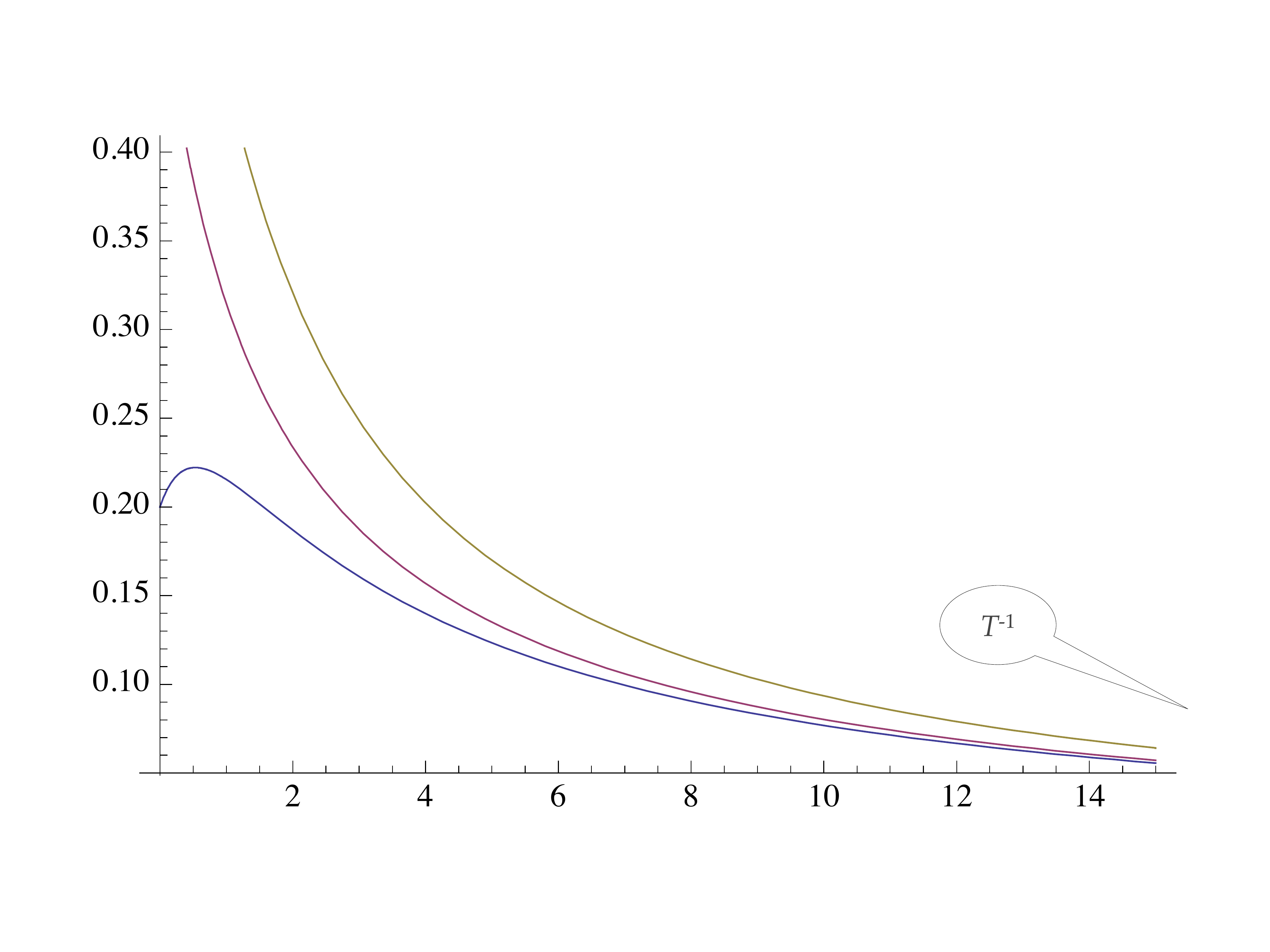}
\end{center}
\caption{ Generic behaviour for
$0<\Omega^1_0<\Omega^2_0<\Omega^3_0$.} \label{halphanisR}
\end{figure}

Solutions (\ref{alleq-R}) and (\ref{2eq-R}) show that the asymptotic behaviour of $\Omega$s is clearly $\nicefrac{1}{T}$, when at least two $\Omega$s are equal. In the more general case, the large-$T$ behaviour is readily obtained thanks to the quasimodular properties of the solutions (see footnote \ref{modtran}):
\begin{equation}    \label{quasimodprop}
 \Omega^{1,2,3}(T)=-\frac{1}{T^2}\Omega^{2,1,3}\left(\frac{1}{T}\right)+ \frac{1}{T}.
     \end{equation}
 Therefore, for  \emph{finite and positive} $\Omega^i_0\equiv \Omega^i(0)$,
\begin{equation}        \label{genbeh}
\Omega^i = \frac{1}{T} + \mathrm{subleading \ at \ large}\ T,
     \end{equation}
 as one observes in Fig. \ref{halphanisR}.    
 Note that this \emph{does not} hold for the solution (\ref{halphsolR}) because for the latter $T=0$ is a pole and $\Omega^i_0\equiv \Omega^i(0)$ is neither finite, nor positive for all $i$.  The behaviour at large $T$ is not $\nicefrac{1}{T}$, but exponential (see Eq. (\ref{lTHo}) and Fig. \ref{halphorig}). 

 As a consequence of the generic behaviour (\ref{genbeh}) of $\Omega$s for positive and finite initial conditions, the late-time geometry on the $S^3$ under the Ricci flow is
\begin{equation}   
\mathrm{d}s^2
 \approx \frac{1}{\sqrt{T}}
\left(\left(\sigma^1\right)^2+
\left(\sigma^2\right)^2+\left(\sigma^3\right)^2\right).
     \end{equation}
This is an isotropic (round) three-sphere of shrinking radius\footnote{At large times, the original $SU(2)$ or $SU(2)\times U(1)$ isometry group gets enhanced to $SU(2)\times SU(2)$, while the volume shrinks to zero. These are generic properties along the Ricci flow: the isometry groups may grow in limiting situations, whereas the volume is never preserved, but shrinks for positive-curvature geometries:
$$
 \frac{\mathrm{d}V}{\mathrm{d}t}
=\frac{1}{2} \int \mathrm{d}^Dx \sqrt{\det g}    g^{ij}
\frac{\partial g_{ij}}{\partial t} =-\frac{1}{2} \int \mathrm{d}^Dx \sqrt{\det g} R.
$$}. It is worth stressing that this universal behaviour is specifically due to the quasimodular properties of the solution, reflected in the non-covariant $\nicefrac{1}{T}$ term of (\ref{quasimodprop}).
 
\section{Bianchi IX foliations and conformal self-duality}

So far, we have considered self-dual solutions of Einstein's equations. These satisfy Eqs. (\ref{sdsec}) and are Ricci flat. Solutions of the Darboux--Halphen system involving quasimodular forms are relevant in particular when Bianchi IX foliations are considered.  
The Lagrange and Darboux--Halphen systems, and more general modular and quasimodular forms emerge, however, in set ups where no self-duality and/or Bianchi IX foliation is assumed. Einstein conformally (anti-)self-dual  spaces \emph{i.e.} quaternionic spaces  turn out to exhibit such  interesting relationships. 

Conformally self-dual Einstein spaces satisfy (see Eqs. (\ref{csd-3})) 
\begin{equation}      \label{asdw}
 \widehat{ \mathcal{W}}^-_i=0.
\end{equation}      
This two--form is defined in (\ref{oswta}) as the anti-self-dual part of the on-shell Weyl tensor (\ref{oswt}). The latter includes a cosmological constant $\Lambda$ and (\ref{asdw}) implies that this space is Einstein ($R_{ab}=\Lambda g_{ab}$) on top of being conformally self-dual ($W^-=0$).

 \subsection{Conformally self-dual Bianchi IX foliations}

Assuming the four-dimensional space be a foliation $\mathcal{M}_4 = \mathbb{R}\times\mathcal{M}_3$ with $\mathcal{M}_3$ a general homogeneous three-sphere invariant under $SU(2)$ isometry, we can in general endow it with a metric (\ref{metans}). The Levi--Civita connection one--forms of the latter are given in (\ref{IX-sdcon}) and (\ref{IX-asdcon})).
Conformal self-duality condition (\ref{asdw}) does not require the flatness of the anti-self-dual component of the connection $A_i$ as in  (\ref{sd-sec}). Hence, no first integral like 
(\ref{flsdLC}) is available. 

In order to take advantage of the conformal self-duality condition (\ref{asdw}) and reach first-order differential equations as in the case of pure self-duality, we can parameterize the connection $A_i$ and $\Sigma_i$ and demand that (\ref{asdw}) be satisfied. This is usually done by setting both $A_i$ and $\Sigma_i$ proportional to $\sigma^i$, as in (\ref{flsdLC}), with a $T$-dependent coefficient though (see e.g. \cite{hep-th/9810005}):
   \begin{eqnarray}
  \Sigma_i &=& -\frac{B_i}{2\Omega^i}\sigma^i,\\ 
  A_i &=& \frac{\Delta_i}{2\Omega^i}\sigma^i.
   \end{eqnarray}
Using Eqs.  (\ref{IX-sdcon}) and (\ref{IX-asdcon}), one obtains a relationship between $\left\{\dot\Omega^i\right\}$ and $\left\{ B_i,\Delta_i\right\}$:
\begin{equation}      \label{BD}
  \dot\Omega^i={\Omega^j\Omega^k}-\Omega^i\left(\Delta_j+\Delta_k\right)= -{\Omega^j\Omega^k}+\Omega^i\left(B_j+B_k\right).
  \end{equation}      
Furthermore, Eqs. (\ref{oswts}) and (\ref{oswta}) lead to the following expressions for the on-shell Weyl tensor:
\begin{eqnarray}
  \widehat{ \mathcal{W}}^+_i&=&-\left\{\frac{1}{4\Omega^1\Omega^2\Omega^3}
  \left(\dot B_i+
      B_jB_k-B_i\left(B_j+B_k\right)\right)+\frac{\Lambda}{6}\right\}\phi^i\nonumber
       \\ &&-\left\{\frac{1}{4\Omega^1\Omega^2\Omega^3}
      \left(\dot B_i-
     B_jB_k-B_i\left(B_j+B_k\right)\right)+\frac{B_i}{2\left(\Omega^i\right)^2}\right\}\chi^i,\label{oswtsIX}\\
          \widehat{ \mathcal{W}}^-_i&=& \left\{\frac{1}{4\Omega^1\Omega^2\Omega^3}
   \left(\dot\Delta_i+
      \Delta_j\Delta_k+\Delta_i\left(\Delta_j+\Delta_k\right)\right)-\frac{\Delta_i}{2\left(\Omega^i\right)^2}\right\}\phi^i\nonumber
       \\  &&+ \left\{\frac{1}{4\Omega^1\Omega^2\Omega^3}
      \left(\dot\Delta_i-
      \Delta_j\Delta_k+\Delta_i\left(\Delta_j+\Delta_k\right)\right)-\frac{\Lambda}{6}\right\}\chi^i. \label{oswtsIXb}
          \end{eqnarray}
The additional (with respect to (\ref{BD})) first-order equations for  $\left\{ B_i\right\}$ or $\left\{\Delta_i\right\}$ are obtained by imposing on-shell conformal self-duality. The canonical method for that is to demand that both  coefficients of $\phi^i$ and $\chi^i$ in (\ref{oswtsIXb}) vanish. This guarantees (see (\ref{oswts})) a conformally self-dual, Einstein manifold with scalar curvature $R=4\Lambda$, in other words a quaternionic space.

Solving the system of equations obtained for conformally self-dual, Einstein manifolds depends drastically on whether or not the isometry is strictly $SU(2)$, \emph{i.e.} the leaves of the Bianchi IX foliation are anisotropic, triaxial spheres. When the isometry is extended to  $SU(2)\times U(1)$ (two equal $\Omega$s), the equations are algebraically integrable (as in the Darboux--Halphen system (\ref{DH})) and no relationship appears with modular or quasimodular forms. This leads to a variety of well known biaxial solutions (see \cite{GP78,P85, hep-th/9810005} as well as \cite{hep-th/0206151} for a detailed presentation of the resolution) such as  (anti-)de Sitter--Taub--NUT, (anti-)de Sitter--Eguchi--Hanson, 
(pseudo-)Fubini--Study -- $\mathbb{C}P_2$, Pedersen (the parentheses correspond to negative $\Lambda$) \dots When all  $\Omega$s are equal, the leaves are round, uniaxial three-spheres, and the only four-geometries are the symmetric $S^4$ or $H_4$ (depending again on the sign of $\Lambda$). 

Although straightforward, the above approach leads for the triaxial case to equations which are not known to be integrable. Hence, their resolution is not systematic and general. An alternative 
strategy has been proposed by Tod and Hitchin \cite{Tod94, Hitchin95}, based on twistor spaces  and isomonodromic deformations (see also \cite{Ward80,LeBrun82}). In a first step, one sets 
 $\Lambda$  to zero in (\ref{oswtsIXb}) and demands the coefficient of $\chi^i$ to vanish. This is equivalent to demanding conformal self-duality and zero scalar curvature ($W^-=s=0$) without setting $C^-_{ij}$ to zero. Thus, the space is  not Einstein and has zero scalar curvature. The final step is to perform a conformal transformation, which allows to restore a non-vanishing scalar curvature, while simultaneously setting $C^-_{ij}=0$. One thus obtains a quaternionic space.  

Explaining the details of this procedure is beyond our present scope, and we will therefore limit our presentation to the issues involving modular forms, which stem out of conditions $W^-=s=0$. These are imposed by demanding that the coefficient of $\chi^i$  vanishes in (\ref{oswtsIXb}) and  setting $\Lambda=0$:
\begin{equation}      \label{sI}
  \mathrm{I}\quad
\begin{cases}
    \dot{\Delta}_1 = \Delta_2 \Delta_3 - \Delta_1 \left(\Delta_2
      + \Delta_3 \right)   \\
    \dot{\Delta}_2 = \Delta_3 \Delta_1 - \Delta_2 \left(\Delta_3
      + \Delta_1 \right)   \\
    \dot{\Delta}_3 = \Delta_1 \Delta_2 - \Delta_3 \left(\Delta_1 +
      \Delta_2 \right).
  \end{cases}
  \end{equation}      
 They are supplemented with Eqs. (\ref{BD}), which read for $\left\{\Delta_i\right\}$:
\begin{equation}       \label{sII} 
  \mathrm{II}\quad
\begin{cases}
    \dot{\Omega}^1 = \Omega^2 \Omega^3 - \Omega^1 \left(\Delta_2
      + \Delta_3 \right)   \\
    \dot{\Omega}^2 = \Omega^3 \Omega^1 - \Omega^2 \left(\Delta_3
      + \Delta_1 \right)   \\
    \dot{\Omega}^3 = \Omega^1 \Omega^2 - \Omega^3 \left(\Delta_1 +
      \Delta_2 \right).
  \end{cases}
  \end{equation}      

Before pursuing the present investigation any further, it is worth making contact with the results of Sec. \ref{foe} on genuine self-duality equations. Assuming the system I and II satisfied \emph{i.e.}  $W^-=s=0$, $\mathcal{A}_i$ (Eq. (\ref{acurv})) reads:
\begin{equation}  
 \mathcal{A}_i =\frac{1}{2}C^-_{ij}\phi^j = \frac{1}{2\Omega^i}\left(
  \frac{ \Delta_j\Delta_k}{\Omega^j\Omega^k}-\frac{\Delta_i}{\Omega^i}
  \right)\phi^i.
  \end{equation}      
Purely self-dual Einstein vacuum spaces are obtained by demanding $C^-_{ij}=0$ (\emph{i.e.} $R_{ab} = 0$ since the scalar curvature vanishes). This leads to the two known possibilities for Bianchi IX vacuum self-dual Einstein geometries met in Sec. \ref{foe}, and satisfying either one of the following systems:
  \begin{itemize}
\item Lagrange (\ref{L}) for $\Delta_i=0$, 
\item Darboux--Halphen  (\ref{DH}) for $\Delta_i=\Omega^i$.
\end{itemize}

  \subsection{Solving I \& II with Painlev\'e VI}

Systems I and II (Eqs. (\ref{sI}) and (\ref{sII})) describing general conformally self-dual Bianchi IX foliations with vanishing scalar curvature ($W^-=s=0$) were studied e.g. in \cite{Pedepoon90, Tod90} prior to their uplift to quaternionic spaces. Further developments in relation with modular properties can be found in \cite{Mas94, Babich:1998tz}.
 
 As usual it is convenient to move to the complex plane, introduce  $\omega^\ell(z)$ and $\delta_\ell(z)$ and trade the dot for a prime as derivative with respect to $z$ in (\ref{sI}) and (\ref{sII}). Real solutions are recovered as previously: $\Omega^\ell (T) = i  \omega^\ell(iT)$ and $\Delta_\ell (T) = i  \delta_\ell(iT)$. 
 
 The system I is that of Darboux--Halphen for $\delta_i(z)$ (see (\ref{DH})). Given a solution $\delta_i(z)$ one can solve the system II  for $\omega^i(z)$. Furthermore,
the $SL(2,\mathbb{C})$ solution-generating technique described in (\ref{slact})
can be generalized in the present case: given  a solution $\delta_i(z)$ and
 $\omega^i(z)$, 
 \begin{equation}       \label{sl2gende}       
      \tilde \delta_i (z) = \frac{1}{\left(c z + d  \right)^2} \delta_i
    \left(\frac{a z + b}{c z + d} \right) + \frac{c}{c z +
    d}
    \end{equation}      
and   
\begin{equation}       \label{sl2genom}
     \tilde \omega^i (z) = \frac{1}{\left(c z + d  \right)^2} \omega^i
    \left( \frac{a z + b}{c z + d} \right) 
    \end{equation}     
provide \emph{another} solution  if         
$\left(\begin{smallmatrix}
      a & b \\ c & d
    \end{smallmatrix}
  \right) \in SL(2, \mathbb{C} )$.
       
    Assuming $\delta_1\neq\delta_2\neq\delta_3$ \emph{i.e.} the triaxial situation (implying automatically $ \omega^1\neq\omega^2\neq\omega^3$), we can readily obtain the general solution of the system I
as in (\ref{DH-sol}),
  \begin{equation}       \label{delsol}      
\delta_i (z)=
    - \frac{1}{2} \frac{\mathrm{d }}{\mathrm{d }z } \log \mathcal{E}^i (z) ,
    \end{equation}     
with $\mathcal{E}^i (z)$ a triplet of weight-two modular forms of
    $\Gamma(2)\subset SL(2, \mathbb{Z} )$. These can be expressed as in  (\ref{schsol}), where  
$\lambda$ is a solution of Schwartz's equation (\ref{sch}). Define now a new set of functions $w_i(z)$ as
  \begin{equation}  
w_i= \frac{\omega^i}{\sqrt{ \mathcal{E}^j \mathcal{E}^k}} 
    \end{equation}           
($i,j,k$ cyclic permutation of $1,2,3$), and insert the solutions (\ref{delsol}) in system II (\ref{sII}). The latter becomes
  \begin{equation}       \label{sIIv2}      
\frac{\mathrm{d}w_1}{\mathrm{d}\lambda}= \frac{w_2w_3}{\lambda}, 
 \quad     \frac{\mathrm{d}w_2}{\mathrm{d}\lambda}= \frac{w_3w_1}{\lambda-1},    \quad     \frac{\mathrm{d}w_3}{\mathrm{d}\lambda}= \frac{w_1w_2}{\lambda(\lambda-1)}.
    \end{equation}           
Notice the first integral $w_1^2-w_2^2+w_3^2$. Even though the value of this integral  is arbitrary, the uplift of the corresponding conformally self-dual geometry with zero scalar curvature to an Einstein manifold is possble only if the constant is $\nicefrac{1}{4}$ (see
\cite{Tod94,Hitchin95}). 

The system of  equations (\ref{sIIv2}) can be solved in full generality with $w_i$ expressed in terms of solutions $y(\lambda)$ of Painlev\'e VI  equation \cite{JM81} (see also \cite{AbCla91} for a more general overview):
     \begin{eqnarray}
  w_1^2&=& \frac{(y-\lambda)y^2(y-1)}{\lambda}
    \left(v-\frac{1}{2(y-1)}\right) \left(v-\frac{1}{2(y-\lambda)}\right),\label{w1pIV}
   \\ 
    w_2^2&=& \frac{(y-\lambda)y(y-1)^2}{\lambda-1},
      \left(v-\frac{1}{2y}\right) \left(v-\frac{1}{2(y-\lambda)}\right)\label{w2pIV}
   \\ 
    w_3^2&=& \frac{(y-\lambda)^2y(y-1)}{\lambda(\lambda-1)}
   \left(v-\frac{1}{2y}\right) \left(v-\frac{1}{2(y-1)}\right).\label{w3pIV}
             \end{eqnarray}
Here,  
        \begin{equation}    
 v=\frac{\lambda(\lambda-1)y'}{2y(y-1)(y-\lambda)}
+ \frac{1}{4y}+ \frac{1}{4(y-1)}- \frac{1}{4(y-\lambda)}
      \end{equation}           
 and 
   $y$ is a solution of Painlev\'e VI equation ($f'=\nicefrac{\mathrm{d}f}{\mathrm{d}\lambda}$): 
   \begin{eqnarray}
  y''&=& \frac{1}{2}\left( \frac{1}{y} +\frac{1}{y-1}+ \frac{1}{y-\lambda}\right)(y')^2-
  \left( \frac{1}{\lambda} +\frac{1}{\lambda-1}+ \frac{1}{y-\lambda}\right)y'  \nonumber   \\  
    &&+ \frac{(y-\lambda)y(y-1)^2}{8\lambda^2(\lambda-1)^2}
      \left({1} -\frac{\lambda}{y^2}
       +\frac{\lambda-1}{(y-1)^1}
       -\frac{3\lambda(\lambda-1)}{(y-\lambda)^2}
      \right).
             \end{eqnarray}
   
 \subsection{Back to quasimodular forms}
 
 We will for concreteness concentrate on the original solution of system I, the Halphen solution corresponding to  $\lambda_{\mathrm{H}} = \nicefrac{\vartheta_2^4}{\vartheta_3^4}$. This is sufficient as any other can be generated by $SL(2,\mathbb{C})$ transformations.  Equations (\ref{sIIv2}) (system II) read now
           \begin{equation}   
   w_1'=i\pi \vartheta_4^4 w_2 w_3,
 \quad     w_2'=-i\pi \vartheta_2^4 w_3 w_1,
 \quad   w_1'=-i\pi \vartheta_3^4 w_2 w_3.
     \end{equation}           
In this form, the system can be solved in terms of Jacobi theta
functions with characteristics \cite{Babich:1998tz}, as an alternative to the solution (\ref{w1pIV})--(\ref{w3pIV}). This makes it relevant in the present framework.

The solution with $ w_1^2-w_2^2+w_3^2=\nicefrac{1}{4}$ -- required for the subsequent promotion to quaternionic geometries -- read:
\begin{eqnarray} 
w_1(z)&= & \frac{1}{2\pi\vartheta_2(0\vert z)\vartheta_3(0\vert z)}\frac{\partial_v  \vartheta \oao{a+1}{b} (0\vert z)}{\vartheta \oao{a}{b} (0\vert z)} ,\label{w1H}
     \\ 
   w_2(z)&= & \frac{\mathrm{e}^{-i\pi \nicefrac{a}{2}}}{2\pi\vartheta_3(0\vert z)\vartheta_4(0\vert z)}\frac{\partial_v  \vartheta \oao{a}{b+1} (0\vert z)}{\vartheta \oao{a}{b} (0\vert z)} ,\label{w2H}
     \\ 
    w_3(z)&= & \frac{-\mathrm{e}^{-i\pi \nicefrac{a}{2}}}{2\pi\vartheta_2(0\vert z)\vartheta_4(0\vert z)}\frac{\partial_v  \vartheta \oao{a+1}{b+1} (0\vert z)}{\vartheta \oao{a}{b} (0\vert z)} .\label{w3H}
\end{eqnarray}
Here $a,b\in \mathbb{C}$ are moduli, mapped under the $SL(2,\mathbb{C})$ transformations (\ref{sl2gende}) and (\ref{sl2genom})  to other complex numbers. If $a,b$ are integers and the transformation is in  $SL(2,\mathbb{Z})$, the solution is left invariant, up to permutation of the three components. 

It would be interesting to present the geometrical structure of the conformally self-dual zero-curvature spaces obtained with the solutions at hand, following the general procedure used in Sec. \ref{geombDH}.  
This would definitely bring us far from the original goal. The interested reader can find useful information in the already quoted literature, both for these spaces and for their quaternionic uplift. Note in that respect that even though many families of solutions exist (here in the triaxial case, or more generally for biaxial three-sphere foliations), very few are singularity-free among which, the Fubini--Study or the  Pedersen instanton ($SU(2)\times U(1)$ isometry), or the Hitchin--Tod solution (strict $SU(2)$ symmetry).

As a final remark, let us mention that (\ref{w1H}), (\ref{w2H}),
(\ref{w3H}) also capture the self-dual Ricci flat solutions  discussed
in Sec. \ref{sdBIX} and given in Eqs. (\ref{halphsol}) \emph{i.e. }the
Atiyah--Hitchin gravitational instanton. They correspond to the choice
$a=b=1\ \mathrm{mod} \ 2 $, that must be implemented with care:
consider $a=1+2\epsilon, b=1+2z_0 \epsilon$ and take the limit
$\epsilon\to 0$. One finds (a useful identity for this computation is
given in
\eqref{e:useful}:
\begin{eqnarray}  
w_1&= & -\frac{1}{\pi\vartheta_2^2\vartheta_3^2}
\left(\frac{i}{z+z_0}-\frac{\pi}{6}
\left(E_2-\vartheta_2^4-\vartheta_3^4
\right)
\right),
     \\  
   w_2&= & -\frac{i}{\pi\vartheta_3^2\vartheta_4^2}
\left(\frac{i}{z+z_0}-\frac{\pi}{6}
\left(E_2+\vartheta_3^4+\vartheta_4^4
\right)
\right), 
     \\  
    w_3&= & -\frac{i}{\pi\vartheta_2^2\vartheta_4^2}
\left(\frac{i}{z+z_0}-\frac{\pi}{6}
\left(E_2+\vartheta_2^4-\vartheta_4^4
\right)
\right) .
\end{eqnarray}
A modulus $z_0$ is left in the solution; under 
 $SL(2,\mathbb{Z})$ it transforms as
\begin{equation} 
z_0 \to \frac{d z_0+b}{c z_0+ a}.
\end{equation}  
For finite $z_0$ the corresponding metric is Weyl-self-dual with zero scalar curvature 
The $z_0\to i \infty$ limit corresponds to the Ricci-flat, Atiyah--Hitchin instanton (Riemann-self-dual). 
 
 \subsection{Beyond Bianchi IX foliations}
\label{sec:beyond}
We would like to close our overview on conformally self-dual geometries with another family of quaternionic solutions, related to modular forms but not of the type $\mathcal{M}_4 = \mathbb{R}\times\mathcal{ M}_3$ with homogeneous $\mathcal{ M}_3$. Indeed, self-duality (Eq. (\ref{sdsec})) or conformal self-duality (Eqs. (\ref{csd-1}) and (\ref{csd-2})) can be demanded outside ot the framework of foliations. 

On can indeed assume an ansatz for the metric of the Gibbons--Hawking type  \cite{Gibbons:1979zt}:
  \begin{equation}   
       \mathrm{d}s^2=\Phi^{-1}\left(\mathrm{d}\tau +\varpi_i \mathrm{d}x^i\right)^2 +\Phi \delta_{ij}\mathrm{d}x^i \mathrm{d}x^j.
  \end{equation} 
Here  $\Phi$ and $\varpi_i$ depend on $\mathbf{x}$ only, and thus $\partial_{\tau}$ is Killing. With this ansatz more general self-dual solutions are obtained with $U(1), U(1)\times U(1)$ or $U(1)\times \mathrm{Bianchi}$  isometry.
Determining quaternionic spaces, \emph{i.e.} conformally self-dual and Einstein, is however far more difficult. It is a real \emph{tour de force} to  find the most general quaternionic solution with $U(1)\times U(1)$ isometry and this was achieved by Calderbank and Pedersen in \cite{CP02}, following  the original method of Lebrun \cite{LeBrun82}. This will be our last example, where a new kind of modular forms emerge.

In coordinates $\{\rho, \eta,\theta,\psi\}$ with frame
  \begin{equation}      
  \alpha=\sqrt{\rho} \mathrm{d}\rho, \quad \beta=\frac{\mathrm{d}\psi +\eta\mathrm{d}\theta}{\sqrt{\rho}}, \quad \gamma= \mathrm{d}\rho, \quad \delta=\mathrm{d}\eta,
  \end{equation} 
  The metric reads:
    \begin{eqnarray}\label{CPU12}
  \mathrm{d}s^2 &=& \frac{4\rho^2\left(F^2_\rho + F^2_\eta \right)-F^2}{4F^2\rho^2}    \left(\gamma^2+\delta^2\right)\nonumber
     \\  &&+ \frac{\left[
     \left(F-2\rho F_\rho\right)\alpha-2\rho F_\eta \beta
     \right]^2+\left[
     \left(F+2\rho F_\rho\right)\beta-2\rho F_\eta \alpha
     \right]^2}{F^2\left[4\rho^2\left(F^2_\rho + F^2_\eta \right)-F^2\right]}.
          \end{eqnarray}
Here $F_\rho=\partial_\rho F$ and $F_\eta=\partial_\eta F$, where   $F(\rho,\eta)$ is a solution of 
  \begin{equation}       \label{harmF}     
\rho^2 \left(\partial^2_\rho+\partial^2_\eta\right)F=\tfrac{3}{4}F.
  \end{equation}     

The metric (\ref{CPU12}) has generically two Killing vectors,  $\partial_{\theta}, \partial_{\psi}$ and $F(\rho, \eta)$ is a harmonic function on $H_2$ with eigenvalue $\nicefrac{3}{4}$. Indeed, the metric on the hyperbolic plane is 
  \begin{equation} 
\mathrm{d}s_{H_2}^2= \frac{ \mathrm{d}\rho^2+ \mathrm{d}\eta^2}{\rho^2}
  \end{equation}     
and Eq. (\ref{harmF})  can be recast as
 \begin{equation}       \label{harmH}     
\triangle_{H_2}F=\tfrac{3}{4}F . 
     \end{equation}     
Solving (\ref{harmH}) leads inevitably to modular forms of $\tau=\eta+i\rho$, even though algebraic solutions are also available. 

Let  us mention for example 
 \begin{equation}  \label{F3h-cp2}
 F=\sqrt{\rho+\frac{\eta^2}{\rho}},
     \end{equation}     
which leads to a metric on $\mathbb{C}P_2$ with $U(1)\times SU(2)$ isometry \cite{P86}, or 
 \begin{equation}       \label{F3h-Hei}     
F=\frac{\rho^2-\rho_0^2}{2\sqrt{\rho}}
 \end{equation}     
  with 
   $U(1)\times \mathrm{Heisenberg}$\footnote{Heisenberg algebra is
     Bianchi II.} symmetry \cite{Ambrosetti:2010tu}. Solutions for
   $F(\rho,\eta)$  with strict  $U(1)\times U(1)$ isometry open
   Pandora's box for non-holomorphic Eisenstein series such as  (see \eqref{maass})
 \begin{equation}   \label{H3h}    
F=E_{\nicefrac{3}{2}}(\tau,\bar \tau),
 \end{equation}     
which has a further discrete  residual symmetry
$SL(2,\mathbb{Z})\subset SL(2,\mathbb{R})$.  These will be discussed
in Sec. \ref{sec:xxx} and we refer to the appendix for some precise definitions.  Very little is known at present on the
geometrical properties of the corresponding quaternionic spaces, or on
the fields of application these spaces could find in physics. In string theory, they are known to describe the moduli space of hypermultiplets in
compactifications on Calabi--Yau threefolds~\cite{Ferrara:1989ik}. The
relevance of the Calderbank and Pedersen metrics in this context was recognized in~\cite{Antoniadis:2003sw}.
For further considerations on the role of modular and quasimodular foms  as string instantonic contributions to the moduli spaces of these compactifications, we refer
to~\cite{Alexandrov:2008ds,Alexandrov:2008nk,Bao:2009fg,Bao:2010cc,Alexandrov:2010np,Alexandrov:2009vj} and in particular to the recent review~\cite{Alexandrov:2011va}.

\section{Beyond the  world of Eisenstein series }
\label{sec:xxx}

To end up this review we would like to elaborate on some connections
between  quantum field theory and modular forms. 
This originates from the specific structure of the perturbative
expansions in string and field theory, and calls for develping more
general modular functions than the non-holomorphic Eisenstein series discussed earlier in these notes.

\subsection{The starting point: perturbation theory}

Perturbative expansions in quantum field theory are expressed as 
sums of multidimensional integrals, obtained by
applying Feynman rules or unitarity constraints. These integrals are plagued
by  various divergences that need to be regulated. It was remarked in~\cite{Broadhurst:1996kc,Broadhurst:1996ye}
that the  coefficients of these divergences are given by multiple zeta
values in four dimensions. 
Since this original work, it has become more and more important to further investigate 
the relationship between 
quantum field theory and the structure of multiple zeta values.
This connection has fostered important mathematical results as in   
instance~\cite{Brown:2011ik,Brown:2009ta}, which 
have been reviewed in 
the recent \emph{S\'eminaire Bourbaki} by Pierre Deligne~\cite{Deligne}.

The next observation is that the above mentioned field-theory Feynman integrals arise as certain limits of string-theory integrals defined on higher-genus Riemann surfaces. They are actually obtained  from the boundary of the moduli of  higher-genus  punctured Riemann surfaces. This bridge to string theory sets a handle to the world of modular functions.

There are indeed two motivations for  embedding the analysis into a string
theory framework. The first is of physical nature:  perturbative string theory is
free of ultraviolet divergences, so it provides a well-defined prescription
for regularizing field-theory divergences. In other words, string theory acts as a
specific regularization from which we expect to learn more on 
the fundamental structure of the quantum field theories. The second
motivation is directly related to the topic of this text: string theory is the ideal arena for exploring 
the number theoretic considerations of
quantum field theory and their close connection with  modular forms.


In the present notes we will focus on the case of the tree level and genus one, following
the string analysis in~\cite{Green:1999pv,Green:2008uj} and the mathematical analysis in~\cite{Goncharov}. We will explain in particular that (non-holomorphic) Eisenstein series are not enough for capturing all available information carried by the integrals under consideration. The presentation will be schematic, aiming at conveying a message rather that providing all technical details. For the latter, the interested reader is referred to the quoted literature.

Let us consider the following integral defined on the moduli space of the
genus-$g$ Riemann surface with four marked points:
\begin{equation}\label{e:Ampgen}
  A^{(g)}(s,t,u)= \int_{\mathcal M_g} \mathrm{d}\mu\, \int_{ \Sigma_g}
  \prod_{i=1}^4 \mathrm{d}^2z_i \, \exp\left( \sum_{1\leq i<j\leq 4} 2\alpha'
    k_i\cdot k_j P(z_i,z_j)\right),
\end{equation}
where $\mathcal M_g$ is the moduli space of the closed Riemann surface $\Sigma_g$
of genus $g$. There are four punctures whose positions $z_i$ are
integrated over. We have introduced  $k_1+k_2+k_3+k_4=0$ with $k_i\cdot k_i=0$ representing 
external massless momenta flowing into each puncture. We will also set the Mandelstam variables 
$s=2k_1\cdot k_2=2k_3\cdot k_4$, $t=2k_1\cdot k_4=2k_2\cdot k_3$ and
$u=2k_1\cdot k_3=2k_2\cdot k_4$, obeying $s+t+u=0$ for the massless states at hand.
The physical scale is the inverse tension of the string $\alpha'$. 

The propagator or Green's function $P(z,w)$ is defined on this Riemann
surface by 
\begin{eqnarray}
 0&=& \int_{\Sigma_g} \mathrm{d}^2z \, \sqrt{-g}\, P(z,w),\\  
\partial_z\bar\partial_{\bar z} P(z,w)&=&2\pi \delta^{(2)}(z)-{2\pi
  g_{z\bar z}\over \int_{\Sigma_g} \mathrm{d}^2z \sqrt{-g}},\\ 
\partial_z\bar\partial_{\bar w} P(z,w)&=&-2\pi \delta^{(2)}(z)+\pi
\sum_{I=1}^g \omega_I(z) (\Im m \Omega)^{-1}_{IJ} \omega_J(w),
\end{eqnarray}
where the $\mathrm{d}s^2=g_{z\bar z} \mathrm{d}z\mathrm{d}\bar z$ is the metric on the Riemann,
$\Omega$ the period matrix, and $\omega_I$ with $1\leq I\leq g$ the
first Abelian differentials.


\subsection{Genus zero: the Eisenstein series}

At genus 0, \emph{i.e.} for the Riemann sphere, the propagator is simply given
by 
\begin{equation}\label{e:prop0}
  P^{(0)}(z,w)= \log|z-w|^2  ,
\end{equation}
and the  integral in~\eqref{e:Ampgen} can be evaluated to give
\begin{eqnarray}
A^{(0)}(s,t,u)&=&{1\over \alpha'^3 stu}  \,
{\Gamma\left(1+\alpha's\right)\Gamma\left(1+\alpha't\right)\Gamma\left(1+\alpha'u\right)\over\Gamma\left(1-\alpha's\right)
  \Gamma\left(1-\alpha't\right)\Gamma\left(1-\alpha'u\right)}\nonumber\\ 
&=&{1\over \alpha'^3 stu}\,\exp\left(-\sum_{n=1}^\infty
  {2\zeta(2n+1)\over2n+1} \left[(\alpha's)^n+(\alpha't)^n+(\alpha'u)^n\right]\right).\label{e:Atree-ex}
\end{eqnarray}
The masses of string theory excitations are integer,  quantized  in units
of $\nicefrac{1}{\alpha'}$. It is  therefore expected that the $\alpha'$ expansion
of the string amplitude in~\eqref{e:Atree-ex} is given by multiple
sums over the integers, but it is remarkable that this expansion involves only odd zeta values of
depth one. For $\alpha'\ll1$ a series expansion representation reads:
\begin{equation}\label{e:Atree}
  A^{(0)}(s,t,u)= \sum_{q\geq-1, p\geq0}  c_{(p,q)}\, \sigma_2^p \sigma_3^q,  
\end{equation}
where we have introduced
$\sigma_2=(\alpha's)^2+(\alpha't)^2+(\alpha'u)^2$ and
$\sigma_3=3\alpha'^3stu$.  Since
$\sigma_1=(\alpha's)+(\alpha't)+(\alpha'u)=0$, we immediately see
that all $\sigma_n=(\alpha's)^n+(\alpha't)^n+(\alpha'u)^n$ with $n\geq2$ are given by~\cite{Green:1999pv}
\begin{equation} 
  {\sigma_n\over n}= \sum_{2p+3q=n}  {(p+q-1)!\over p!q!} \, \left(\sigma_2\over2\right)^p \left(\sigma_3\over3\right)^q  .
\end{equation}
The coefficients $c_{(p,q)}$ are polynomial in odd zeta values of
weight $2p+3q-3$. It is notable that at a given order $n=2p+3q-3$, 
the space of these coefficients
has dimension $d_n=\left\lfloor(n+2)/2\right\rfloor-\left\lfloor(n+2)/3\right\rfloor$, which coincides with the dimension of the space of the holomorphic
Eisenstein series of weight $n$ (see appendix). This hint calls for further investigation, and we would like to mention the    
recent work                                                                     
connecting the $\alpha'$ expansion in~\eqref{e:Atree} and the motivic           
multiplet zeta values~\cite{SchSt}.

\begin{figure}[ht]
  \centering
  \includegraphics[width=8cm]{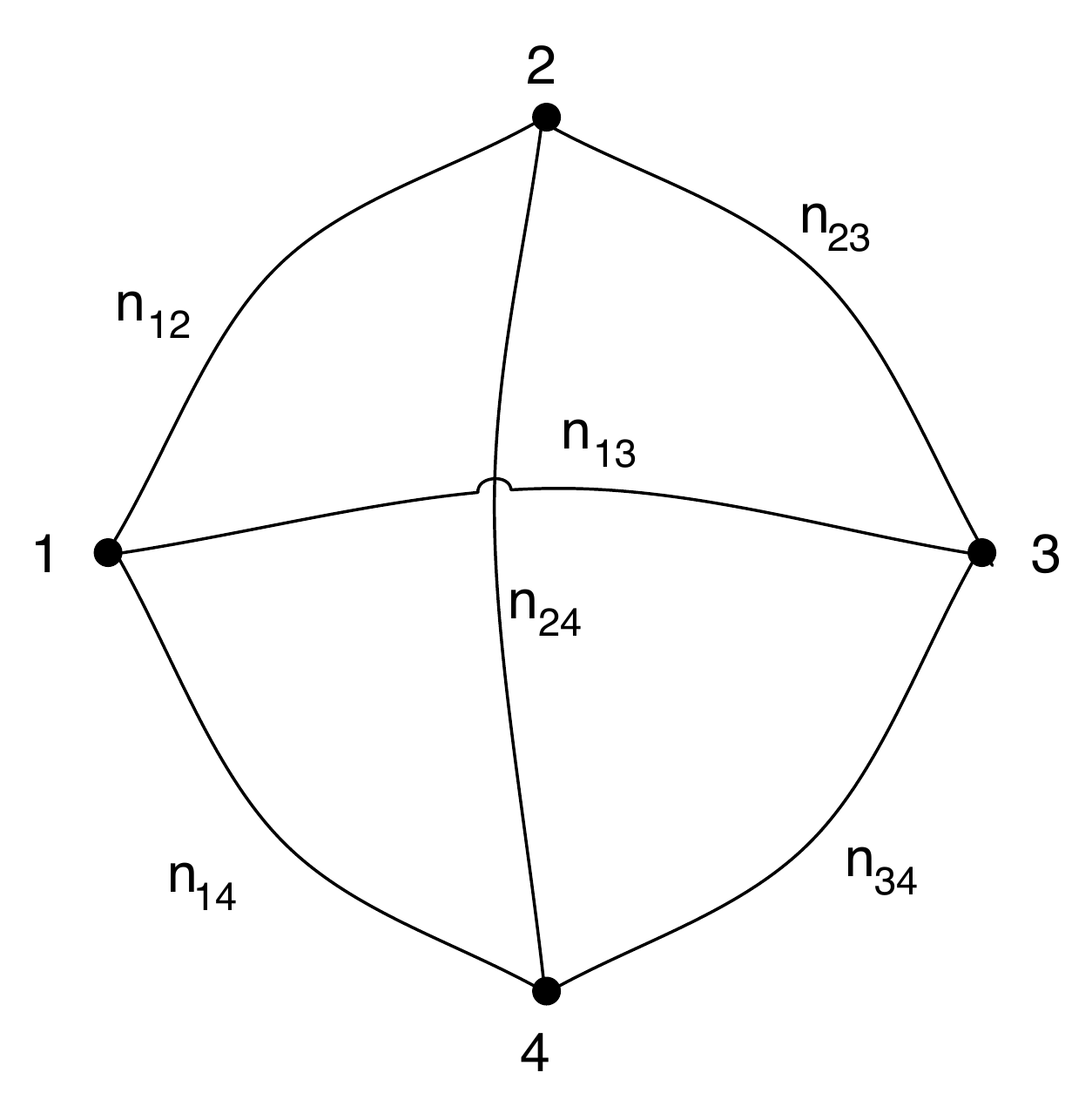}
  \caption{ Graph of a vacuum Feynman diagram on the Riemann surface
    $\Sigma_g$. The punctures are connected by $n_{ij}\geq0$ links
  representing the number of two-dimensional propagators.}
  \label{fig:vacgraphs}
\end{figure}

One can expand the integrand of~(\ref{e:Atree}) and obtain each
coefficient $c_{(p,q)}$ as a linear combination of the  multiple integrals of the propagator
$P^{(0)}(z_i,z_j)$ (given in (\ref{e:prop0})):
\begin{equation}\label{e:intzero}
  c_{n_{12},n_{13},n_{14},n_{23},n_{24}, n_{34}}= \int_{S^2} \prod_{1\leq i<j\leq
    4}\mathrm{d}^2z_i  \, \prod_{1\leq i<j\leq 4} P^{(0)}(z_i,z_j)^{n_{ij}}.
\end{equation}
The integrand of this expression is  the product of the propagators
connecting the punctures  with multiplicities $n_{ij}$, 
$1\leq i<j\leq4$, as depicted on Fig.~\ref{fig:vacgraphs}.
The contributions in~(\ref{e:Atree}) are the
lowest-order to the full string-theory amplitude of
the four-point (four punctures) processes described here.

In Eqs.~(\ref{e:Atree-ex})--(\ref{e:Atree}), we encountered the zeta values 
\begin{equation}\label{e:zv}
  \zeta(s)=\sum_{n\geq 1} {1\over n^s}  .
\end{equation}
Extending the sum over the integers $n$ to a lattice like $\mathbf
p=m+\tau n\in \Lambda^{(1)}=\mathbb Z+\tau\mathbb Z$, where\footnote{The modular parameter $\tau$ is expressed in alternative ways throughout these notes: $\tau=\Re
   e(\tau)+i\Im   m(\tau)=\tau_1+i\tau_2=\eta+i\rho$.} $\tau\in \mathfrak h=\{z\in\mathbb C,
\Im m(z)\geq0\}$, one gets the (non-holomorphic) Eisenstein series
\begin{equation} 
 \hat E_s(\tau, \bar \tau)= \sum_{\mathbf p\in \mathbb Z+\tau\mathbb Z} {1\over |\mathbf p|^{2s}} ,
\end{equation}
where $|\mathbf p|^2=(m+n\tau)(m+n\bar\tau)$ is the natural Euclidean norm  on
the lattice $\Lambda^{(1)}$.
This expression can be made modular-invariant in a trivial way by
multiplying by $\Im m(\tau)^s$, 
\begin{equation}
   \label{e:Es}
 E_s(\tau, \bar \tau)= \sum_{\mathbf p\in \mathbb Z+\tau\mathbb Z} {\Im
   m(\tau)^s\over |\mathbf p|^{2s}} .
\end{equation}
This Eisenstein series is an eigenfunction of the hyperbolic
Laplacian (\ref{harmF})
%
with eigenvalue $s(s-1)$:
\begin{equation}    \label{e:lapl}
\triangle_{H_2} E_s(\tau,\bar\tau) = s(s-1) \, E_s(\tau,\bar\tau).
\end{equation}
The case $s=\nicefrac{3}{2}$ was discussed in Sec.~\ref{sec:beyond}, Eqs. (\ref{F3h-cp2})--(\ref{H3h}), from a different physical perspective.

\subsection{Genus one: beyond}

At this stage of the exposition the reader may wonder how the above generalization of the zeta values (\ref{e:zv}) into modular forms (\ref{e:Es}) arises in  string theory. We will sketch how this goes and show that new automorphic forms are actually needed, standing beyond
the well known Eisenstein series. This requires going beyond the sphere (\ref{e:prop0})--(\ref{e:Atree}).

At genus one, the Green's function is given by 
\begin{equation}\label{e:Pone}
  P^{(1)}(z,0) = -\frac14 \log\left| \vartheta_1(z|\tau)\over
    \vartheta_1'(0|\tau)\right|^2 +  {\pi\Im m (z)^2\over 2\tau_2}.
\end{equation}
The amplitude in~(\ref{e:Ampgen}) reads:
\begin{equation}\label{e:Aone}
  A^{(1)}(s,t,u)= \int_{\mathcal F}{\mathrm{d}^2\tau\over\tau_2^2}
  \int_{\mathcal T}    
  \prod_{1\leq i<j\leq 4} {\mathrm{d}^2z_i\over\tau_2}\,\mathcal W^{(1)} \,
  \mathrm{e}^{-\sum_{1\leq i<j\leq4}2\alpha' k_i\cdot k_j \hat P^{(1)}(z_i-z_j)},
\end{equation}
where $\mathcal F=\left\{\tau; |\Re e(\tau)|\leq \frac12, \Im m(\tau)>0,
\Re e(\tau)^2+\Im m(\tau)^2\geq1\right\}$ is a fundamental domain for $SL(2,\mathbb Z)$, and
$\mathcal T=\left\{z; |\Re e(z)|\leq\frac12, 0\leq \Im m(z)\leq\Im
m(\tau)\right\}$.

No closed form for the integral in~\eqref{e:Aone} is known, in particular because of the
presence of non-analytic contributions in the complex
$(s,t)$-plane. For a rigorous definition of this integral we refer to~\cite{D'Hoker:1994yr}.
The expression for the genus-one propagator in~\eqref{e:Pone} has an alternative representation: 
\begin{equation} 
  P^{(1)}(z,0)={1\over2\pi} \sum_{\mathbf p\in\mathbb Z+\tau\mathbb Z}\,
  {\tau_2\over |\mathbf p|^2}\, \mathrm{e}^{-\pi {\Im m (\bar z \mathbf p)\over\tau_2}} +C(\tau,\bar \tau),
\end{equation}
where $C(\tau,\bar \tau)=\log|\sqrt{2\pi} \eta(\tau)|$ is a modular anomaly. Since the latter is $z$-independent, it drops out of the sum
in~(\ref{e:Aone}) because of the momentum-conservation condition $\sum_{i=1}^4 k_i=0$.  The
integrand of~\eqref{e:Aone} is therefore modular-invariant. 
From now on we will only consider the modular-invariant part of the propagator
\begin{equation}
  \hat P^{(1)}(z,0)={1\over2\pi} \sum_{\mathbf p\in\mathbb Z+\tau\mathbb Z}\,
  {\tau_2\over |\mathbf p|^2}\, \mathrm{e}^{-\pi {\Im m (\bar z \mathbf p)\over\tau_2}} .
\end{equation}

Following the developments on the sphere, we can analyze the expansion of the amplitude (\ref{e:Aone}) for $\alpha'\ll1$. In this regime one gets integrals
of the type~\eqref{e:intzero}, but this time with the genus-one propagator 
\begin{equation} 
  D_{n_{12},n_{13},n_{14},n_{23},n_{24}, n_{34}}(\tau,\bar\tau)= \int_{\mathcal T} \prod_{1\leq i<j\leq
    4}{\mathrm{d}^2z_1  \over \tau_2}\, \prod_{1\leq i<j\leq 4} \hat P^{(1)}(z_i,z_j)^{n_{ij}}.
\end{equation}
The product runs over the entire set of links with multiplicities $n_{ij}$,
$1\leq i<j\leq4$, of the graph $\Gamma$ depicted in Fig~\ref{fig:vacgraphs}.
By construction these integrals are modular functions for $SL(2,\mathbb
Z)$.
Performing the integration over the position of the punctures one gets
an alternative form for the modular function $D_{n_{12},\ldots,n_{34}}(\tau,\bar \tau)$  given
by 
\begin{equation} 
  D_{n_{12},\ldots,n_{34}}(\tau,\bar \tau)  = \sum_{p_i\in \Gamma}\, \prod_{i=1}^4\delta\left(\sum_{j\to
    v_i} \mathbf p_j\right)  \prod_{\mathrm{prop}\in\Gamma} {\Im m(\tau)\over|\mathbf p_i|^2},
\end{equation}
where the sum is over all the propagators $p_i$ of the graph
$\Gamma$. If there are $n_{12}$ propagators connecting the vertices 1
and 2, we have $n_{12}$ different elements of the lattice
$p_i=m_i+\tau n_i\in \mathbb Z+\tau\mathbb Z$, $1\leq i\leq
n_{12}$. At each vertex $v_i$ of the graph we impose momentum conservation
by demanding that the sum of the incoming momenta $\mathbf p_j$ flowing to this
vertex ($j\to v_i$) be zero. This is represented by the delta function
constraint $\delta(\sum_{j\to
    v_i} \mathbf p_j)$ with
$\delta(m+\tau n)\equiv\delta(m)\delta(n)$.

The above sums $D_{n_{12},\ldots,n_{34}}(\tau,\bar\tau)$, introduced in~\cite{Green:2008uj}, are generalizations of the
Eisenstein series, that we will call \emph{Kronecker--Eisenstein} 
following~\cite{Goncharov}. 
With each modular function $D_{n_{12},\ldots,n_{34}}(\tau,\bar \tau)$  we associate a weight
given by the sum of the integer-valued indices $n_{ij}$.
Let us  focus for concreteness on the particular case of $n$ propagators between two
punctures, and refer to~\cite{Green:2008uj} for the general case. We define 
\begin{equation}
  \label{e:defDn}
  D_{n}(\tau,\bar\tau):= \sum_{\mathbf  p_i\in \mathbb
    Z+\tau\mathbb Z}\,\delta\left(\sum_{i=1}^n \mathbf p_i\right)
 \prod_{i=1}^n {\Im m(\tau)\over 4\pi|\mathbf p_i|} .
\end{equation}
The special cases $n=2$ and $3$ are given\footnote{The case $n=3$ has been worked out by Don Zagier.} in~\cite[appendix~B]{Green:2008uj} 
\begin{eqnarray} 
D_{2}(\tau,\bar\tau)&= &\frac{E_2(\tau,\bar \tau)}{(4\pi)^2},\\
D_3(\tau,\bar\tau)&=&\frac{E_3(\tau,\bar \tau)}{(4\pi)^3}+\frac{\zeta(3)}{64} .
\end{eqnarray}
However, in  general these modular functions do not reduce to Eisenstein series,
as it can easily be seen by evaluating the constant terms. 
For $n\geq2$ it is always possible to decompose the modular form
$D_n(\tau,\bar\tau)$ as~\cite{Green:2008uj}
\begin{equation}\label{e:Ddn}
D_n(\tau,\bar\tau)= P_n(E_s(\tau, \bar \tau))+ \delta_n(\tau,\bar\tau)
\end{equation}
with $P_n(E_s(\tau, \bar \tau))$  a polynomial in the Eisenstein series
$E_s(\tau, \bar \tau)$ (see Eqs.~\eqref{e:Es} and \eqref{maass}) of the form
\begin{equation} 
  P_n(E_s(\tau, \bar \tau))=p_n(\zeta(2n+1))+  {b_1\over(4\pi)^n}\, E_n(\tau, \bar \tau)+ \sum_{r+s=n}
  {c_{r,s}\over (4\pi)^n} E_r(\tau, \bar \tau) E_s(\tau, \bar \tau),
\end{equation}
where $p_n(\zeta(2n+1))$ is polynomial of degree two in the odd zeta values of total
weight $n$.
The remainder $\delta_n(\tau,\bar \tau)$ in~\eqref{e:Ddn} is a modular form whose 
constant Fourier coefficient does not  vanish but tends to zero for
$\tau_2\to\infty$.

Although the definition of the modular functions $D_{n_{12},\ldots,n_{34}}(\tau,\bar\tau)$ given in~\eqref{e:defDn} looks similar to the double-Eisenstein series introduced
in~\cite{ZagierOhnoGangl}, one finds that, as opposed to the latter, their constant term 
involves depth-one
zeta values~\cite{Green:2008uj} only. Hence, they provide a natural modular-invariant generalization of the polynomials in the odd zeta values met in~\eqref{e:intzero}.
One way to obtain multiple zeta values is to insert
in~\eqref{e:defDn}  the generalized
propagator used by  Goncharov in~\cite{Goncharov}. 
Whether the generalization introduced by  Goncharov 
does appear in string theory is an open question. From the original
physical perspective, this question is relevant because it translates
into the (im)possibility of appearance of multiple zeta values as counter-terms to ultraviolet divergences in quantum field theory. This might have important consequences in supergravity. 
 
As a final comment, let us mention that although our discussion was
confined to the case of modular functions for  $SL(2,\mathbb Z)$,
most of the above can be generalized to the framework
of  automorphic functions for  
higher-rank group~\cite{Green:2010kv}.
 
 \section{Concluding remarks}

In the present lecture notes we have given a partial -- in all possible senses -- review of the emergence of (quasi)modular forms in the context of gravitational instantons and string
theory. These forms often appear as the consequence of remarkable, explicit or hidden symmetries, and turn out to be valuable tools for unravelling a great deal of properties in a variety of physical set ups. The latter include monopole scattering, Ricci flows, non-perturbative (instantonic) corrections  to string moduli spaces (via their Fourier coefficients), or perturbative expansions in quantum field theory (via string amplitudes).

We have described how the classical holomorphic Eisenstein series,
whose theory is nicely presented in~\cite{serre,Koblitz}, occurs in the context of
gravitational instantons or in studying non-perturbative effects.   We have also encountered the
non-holomorphic  Eisenstein series, the analytic properties of which are
described in~\cite{langlands,MW}.  
This whole analysis has led us to argue that one needs novel types of
modular functions, standing beyond the usual Eisenstein series. Although the analytic
properties of these series are still poorly understood, they seem
to be a corner stone for understanding the challenging nature of interactions in
string theory and its consequences in quantum field theory.

 \section*{Acknowledgements}
The content of these notes was presented in 2010 at the Besse summer school on quasimodular forms by P.M. Petropoulos, who is grateful to the organizers and acknowledges financial support by the ANR programme MODUNOMBRES and the \emph{Universit\'e Blaise Pascal}. He would like to thank also the \emph{University of Patras} for kind hospitality at various stages of preparation of the present work. The authors benefited from discussions with G. Bossard, J.--P. Derendinger, A. Hanany, H. Nicolai, N. Prezas, K. Sfetsos and P. Tripathy. The material  is 
based, among others, on works made in collaboration with I. Bakas,
F. Bourliot, J. Estes, M.B. Green, D. L\"ust, S.D. Miller, D. Orlando,
V. Pozzoli, J.G. Russo, K. Siampos,  Ph. Spindel and D. Zagier. This research was supported by the LABEX P2IO, the ANR contract  05-BLAN-NT09-573739, the PEPS-2010 contract BFC-68788, the ERC Advanced Grant  226371,
the ITN programme PITN-GA-2009-237920 
and the IFCPAR CEFIPRA programme 4104-2.

\appendix
\section{Theta functions and Eisenstein series}
\label{sec:theta}

We collect here some conventions for the modular forms and theta functions
used in the main text. General results and properties of these objects can be found in~\cite{serre,Koblitz}.

Introducing  $q=\exp 2i\pi z$, we first define 
\begin{eqnarray}\label{A1}
\eta(z)
&=&q^{\nicefrac{1}{24}}
\prod_{n=1}^\infty\left(1-q^n\right),\\
E_2(z)&=&\frac{12}{i\pi}\frac{\mathrm{d}}{\mathrm{d}z}
\log \eta,
\end{eqnarray}
 as the Dedekind function and the weight-two quasimodular form, whereas
 \begin{equation}
  \vartheta_2(z)=\sum_{p\in \mathbb{Z}} q^{\nicefrac{1}{2}\left(
  p+\nicefrac{1}{2}\right)^2},\quad 
  \vartheta_3(z)=\sum_{p\in \mathbb{Z}} q^{\nicefrac{p^2}{2}},\quad
  \vartheta_4(z)=\sum_{p\in \mathbb{Z}} (-1)^p\, q^{\nicefrac{p^2}{2}}
\end{equation}
are the Jacobi theta functions. More generally, one introduces
\begin{equation}
 \vartheta \oao{a}{b} (v\vert z)= \sum_{m\in\mathbb{Z}}
 \exp\left(i\pi z(m+\nicefrac{a}{2})^2+2i\pi (v+\nicefrac{b}{2})(m+\nicefrac{a}{2})\right)
\end{equation}
with
$\vartheta \oao{1}{1}=  \vartheta_1, \vartheta \oao{1}{0}=
\vartheta_2, \vartheta \oao{0}{0}=  \vartheta_3, \vartheta \oao{0}{1}=
\vartheta_4$. Let us also mention the following relation 
\begin{equation}\label{e:useful}
\vartheta \oao{\alpha + 2 w}{\beta +  2 v} (0\vert z)= \vartheta \oao{\alpha + 2 w}{\beta } (v\vert z)=
\mathrm{e}^{i\pi w(w+ 1+2v)} \vartheta \oao{\alpha}{\beta} (v+wz\vert z).
\end{equation}

The first holomorphic Eisenstein series are
\begin{equation}
\begin{cases}
   E_2(z) = 1-24\sum_{m=1}^\infty\frac{mq^m}{1-q^m}\\ \label{A6}
   E_4(z) = 1+240\sum_{m=1}^\infty\frac{m^3q^m}{1-q^m}\\ 
   E_6(z) =1-504\sum_{m=1}^\infty\frac{m^5q^m}{1-q^m}.\end{cases}
\end{equation}
Notice that $E_4(z)$ and $E_6(z)$ are modular forms of weight 4 and 6,
  whereas $E_2(z)$ is the already quoted weight-two quasimodular form. The modular-invariant
  of weight two is the non-holomorphic combination $E_2(z)-\nicefrac{3}{\pi\Im m(z)}$. It is a classical result that the space of modular
  forms of weight $k$ is spanned by $E_4^aE_6^b$ with $2a+3b=k$. The
  dimension of this space is
  $d_k=\left\lfloor(k+2)/2\right\rfloor-\left\lfloor(k+2)/3\right\rfloor$.

In the main text we also consider non-holomorphic Eisenstein
series $E_s(z,\bar z)$ with $z=x+iy$, $y>0$ and $x\in\mathbb R$. 
These are defined as modular-invariant eigenfunctions  of the hyperbolic Laplacian (see Eqs. (\ref{harmF}),  \eqref{e:Es} and \eqref{e:lapl})
\begin{equation}
  y^2 (\partial_x^2+\partial_y^2) \, E_s(z,\bar z)  = s(s-1)
  E_s(z,\bar z),
\end{equation}
with polynomial  growth at the cusps ($y\to\infty$):
\begin{equation}
E_s(z,\bar z) = \sum_{(m,n)\neq(0,0)}{y^s\over |mz+n|^{2s}},
\end{equation}
for $s\in \mathbb C$ with large enough real part for convergence. One
can extend the definition by analytic continuation~\cite{langlands} for all $s\neq1$
using the functional equation $\Gamma(s)\pi^{-s}E_s(z,\bar
z)=\Gamma(\nicefrac12-s)\pi^{\nicefrac12-s}\,E_{1-s}(z,\bar z)$. 
These series have the following Fourier expansion: 
 \begin{equation}     \label{maass}
2\xi(2s) E_s(z,\bar z)= 2\xi(2s) \, y^s+ 2\xi(2s-1)\,y^{1-s} 
+4y^{\nicefrac{1}{2}}\sum_{n\neq0}{\sigma_{2s-1}(|n|)\over |n|^{s-\nicefrac12}}
K_{s-\nicefrac{1}{2}}(2\pi\vert n\vert y)\,\mathrm{e}^{2\pi i nx}
 \end{equation}     
where 
$\xi(s)=\zeta(s)\Gamma(\nicefrac{s}{2})\pi^{-\nicefrac{s}{2}}$ is the completed zeta function, $K_{s-\nicefrac{1}{2}}$ is the $K$-Bessel function and $\sigma_\alpha (n)=\sum_{d|n} d^\alpha$  (see e.g. \cite{AT} for details).

Finally, let us mention how the non-holomorphic series are connected to the holomorphic ones. 
For that, one considers the following generalization of the non-holomorphic Eisenstein
functions:
\begin{equation}
  E^{(w,\bar w)}_s(z,\bar z)=\sum_{(m,n)\neq(0,0)} \, {y^{s+w+\bar
      w}\over (mz+n)^{s+w} (m\bar z  +n)^{s+\bar w}}.
\end{equation}
These series transform under a modular transformation  $\gamma=
\left(\begin{smallmatrix}
  a&b\cr c&d
\end{smallmatrix}\right) \in SL(2,\mathbb Z)$ as
\begin{equation}
  E_s^{(w,\bar w)}(\gamma\cdot z,\gamma\cdot\bar z)= (c z+d)^w (c\bar
  z+d)^{\bar w}\, E_s^{(w,\bar w)}(z,\bar z).
\end{equation}
Chosing $s=n\in\mathbb N$ and $\bar w=-n$, we recover the
holomorphic Eisenstein series
$E_n^{(0,-n)}(z,\bar z)=  E_n(z)$.

\end{document}